\newif\iffigs\figstrue
\documentclass[10pt,a4paper]{article}
\usepackage{latexsym,amssymb,lscape,graphics,setspace}
\usepackage{amsmath}
\usepackage{graphicx}        
\usepackage{longtable}
\usepackage{multirow}
\usepackage{booktabs}
\usepackage{multirow}
\usepackage{color}
\usepackage{slashed,epsfig}
\usepackage{amsfonts}
\usepackage{cite}
\usepackage{verbatim}
\usepackage{colortbl}
\usepackage[table]{xcolor}
\usepackage{hyperref}       
\usepackage{array}
\usepackage{fancyhdr}
\usepackage{tikz}
\usepackage{multicol}
\usepackage{lscape}
\usepackage{colordvi}
\graphicspath{{images/}}

\newcommand{\mathsym}[1]{{}}

\newtheorem{definizione}{Definition}[section]

\newcommand{\bd}{\begin{definizione}}
\newcommand{\ed}{\end{definizione}}

\newcommand  {\Rbar} {{\mbox{\rm$\mbox{I}\!\mbox{R}$}}}
\newcommand  {\Hbar} {{\mbox{\rm$\mbox{I}\!\mbox{H}$}}}
\newcommand {\Cbar}
    {\mathord{\setlength{\unitlength}{1em}
     \begin{picture}(0.6,0.7)(-0.1,0)
        \put(-0.1,0){\rm C}
        \thicklines
        \put(0.2,0.05){\line(0,1){0.55}}
     \end {picture}}}
\def\IC{\relax\,\hbox{$\inbar\kern-.3em{\rm C}$}}
\def\IG{\relax\,\hbox{$\inbar\kern-.3em{\rm G}$}}
\def\IB{\relax{\rm I\kern-.18em B}}
\def\ID{\relax{\rm I\kern-.18em D}}
\def\IL{\relax{\rm I\kern-.18em L}}
\def\IF{\relax{\rm I\kern-.18em F}}
\def\IH{\relax{\rm I\kern-.18em H}}
\def\II{\relax{\rm I\kern-.17em I}}
\def\IN{\relax{\rm I\kern-.18em N}}
\def\IP{\relax{\rm I\kern-.18em P}}
\def\IQ{\relax\,\hbox{$\inbar\kern-.3em{\rm Q}$}}
\def\bfzero{\relax\,\hbox{$\inbar\kern-.3em{\rm 0}$}}
\def\IK{\relax{\rm I\kern-.18em K}}
\def\IG{\relax\,\hbox{$\inbar\kern-.3em{\rm G}$}}
 \font\cmss=cmss10 \font\cmsss=cmss10 at 7pt
\def\IR{\relax{\rm I\kern-.18em R}}
\def\ZZ{\relax\ifmmode\mathchoice
{\hbox{\cmss Z\kern-.4em Z}}{\hbox{\cmss Z\kern-.4em Z}}
{\lower.9pt\hbox{\cmsss Z\kern-.4em Z}} {\lower1.2pt\hbox{\cmsss
Z\kern-.4em Z}}\else{\cmss Z\kern-.4em Z}\fi}
\def\bfone{\relax{\rm 1\kern-.35em 1}}

\def\inbar{\vrule height1.5ex width.4pt depth0pt}
\def\bfzero{\relax{\rm I\kern-.18em 0}}
\def\bfone{\relax{\rm 1\kern-.35em 1}}

\def\o#1#2{{{#1}\over{#2}}}
\DeclareFontFamily{U}{rsf}{} \DeclareFontShape{U}{rsf}{m}{n}{
  <5> <6> rsfs5 <7> <8> <9> rsfs7 <10-> rsfs10}{}
\DeclareMathAlphabet\Scr{U}{rsf}{m}{n}

\def\T{T}

\newcommand{\SO}{\mathop{\rm SO}}

\newcommand{\USp}{\mathop{\rm {}USp}}
\newcommand{\Sp}{\mathop{\rm {}Sp}}

\newcommand{\ft}[2]{{\textstyle\frac{#1}{#2}}}

\def\1bar{1\hskip -.275cm -}
\def\2bar{2\hskip -.275cm -}
\def\3bar{3\hskip -.275cm -}

\newsavebox{\uuunit}
\sbox{\uuunit}
{\setlength{\unitlength}{0.825em}
\begin{picture}(0.6,0.7)
\thinlines
\put(0,0){\line(1,0){0.5}}
\put(0.15,0){\line(0,1){0.7}}
\put(0.35,0){\line(0,1){0.8}}
\multiput(0.3,0.8)(-0.04,-0.02){10}{\rule{0.5pt}{0.5pt}}
\end {picture}}

\makeatletter \@addtoreset{equation}{section} \makeatother

\def\bfone{\relax{\rm 1\kern-.35em 1}}

\def\bfone{\relax{\rm 1\kern-.35em 1}}
\font\cmss=cmss10 \font\cmsss=cmss10 at 7pt

\newcommand{\so}{\mathfrak{so}}
\newcommand{\su}{\mathfrak{su}}

\newcommand{\uu}{\mathfrak{u}}
\newcommand{\sym}{\mathfrak{sp}}
\newcommand{\slal}{\mathfrak{sl}}

\newcommand{\rmd}{\mathrm{d}}
\newcommand{\rmi}{\mathrm{i}}

\newcommand{\ir}{r} 
\newcommand{\js}{s} 
\newcommand{\ial}{\alpha } 
\newcommand{\jbe}{\beta }
\newcommand{\kga}{\gamma }
\newcommand{\mua}{a} 
\begin{document}
\begin{titlepage}
\begin{center}
\vskip 0.2cm
{{\large {\sc Homogeneous Non Symmetric Special K\"ahler Geometries}\\
{\sc as
Broken Isometry Metrics on Symmetric CV {K}\"ahler Manifolds:\\}}
\vskip 0.1cm
  {\large {\sc a new tool for $r=2$ CaNNs } ${}^\dagger$}} \\
 \vskip 1cm {\sc Pietro Fr\'e\,$^{a,b}$,
 Mario Trigiante\,$^{c, d}$} \\{\sc and  Antoine Van Proeyen \,$^{e}$} \vskip 0.5cm
\smallskip
{\sl \small \frenchspacing ${}^a\,$ {\tt Emeritus Professor of}  Dipartimento di Fisica,
Universit\`a di Torino, Via P. Giuria 1, I-10125 Torino, Italy \\[2pt]
${}^{b}\,${\tt Senior Consultant of } Additati\&Partners Consulting s.r.l,
Via Filippo Pacini 36, I-51100 Pistoia, Italy \\[2pt]
${}^c\,$Dipartimento DISAT, Politecnico di Torino,
C.so Duca degli Abruzzi 24, I-10129 Torino, Italy\\[2pt]
${}^d\,$INFN, Sezione di Torino\\[2pt]
${}^e\,$ {\tt Emeritus Professor of} KU Leuven, Institute for Theoretical Physics
and Leuven Gravity Institute, KU Leuven,
Celestijnenlaan 200D, box 2415, B-3001 Leuven, Belgium \\[2pt]
E-mail:  {\tt pietro.fre@unito.it, mario.trigiante@polito.it, \\ antoine.vanproeyen@kuleuven.be
 } }
\begin{abstract}
In this paper we prove in full detail the isomorphism between the solvable group $\mathcal{S}_{2,2+p}$, metric equivalent to the Calabi Vesentini symmetric space $\mathrm{SO(2,2+p)/SO(2)\times SO(2+p)}$, and the solvable group $\mathcal{S}_{\mathrm{L}(-1,p)}$ supporting the K\"ahler metric of the homogeneous non symmetric special manifold L$(-1,p)$. From an Alekseevskyan point of view the two spaces simply correspond to two different quadratic forms on the same solvable Lie algebra that we present and compare in detail. Furthermore considering the full group of isometries $\mathrm{Iso}_{\mathrm{L}(-1,p)}$ of L$(-1,p)$ we show that $\mathrm{Iso_{L(-1,p)}}\subset \mathrm{SO}(2,2+p)$ is  a non-semisimple subgroup of the simple isometry group of Calabi-Vesentini manifolds. Altogether the Special K\"ahler manifold L$(-1,p)$ can be seen as a coset manifold $\mathrm{Iso_{L(-1,p)}}/\mathrm{H}$
where $\mathrm{H}=\mathrm{U(1)}_L\times \mathrm{SO(p)}$, the generator of $\mathrm{U(1)}_L$ being in $\so(2,2+p)$, yet not in the canonical $\so(2)\oplus\so(2+p)$ subalgebra.  The action of $\mathrm{Iso}_{\mathrm{L}(-1,p)}$ on the solvable coordinates of the CV manifold can be constructed directly. This identification provides a valuable tool for   Cartan Neural Networks, introducing additional non linear transformations in every map from one layer to the next one of an $r=2$ Neural Network based on the CV Tits Satake universality class; in perspective, this new tool increases expressivity.
\end{abstract}
\vfill
\end{center}
\noindent \parbox{175mm}{\hrulefill}
\par
${}^\dagger$ P.G. Fr\'e acknowledges support by the Company \textit{Additati\&Partners
Consulting s.r.l} during the development of
the present research.\\
\\[5pt]
\end{titlepage}
{\small \tableofcontents} \noindent {}
\newpage
\section{Introduction}
The present paper is the ninth episode in the development of a general research plan, started about three  years ago and first publicly announced March 2025 by means of the foundational paper \cite{pgtstheory}, the previous seven episodes, apart from the quoted foundational one, being the articles \cite{TSnaviga,naviga,tassellandum,axialbeltra,geotermico,secondtemperature,terzatemperatura}. The aim of such a research programme is the geometrical refoundation of Artificial Intelligence on the basis of Cartan Neural Networks (\textbf{CaNNs}),   defined in \cite{TSnaviga,naviga}, (June-July 2025), the long time goal being to establish  new Deep Learning Algorithms that should be free from the \textit{original sin}, namely \textit{the point-wise activation functions} that, together with the inappropriate use of \textit{flat Euclidean manifolds} as models of the layers, constitute the root of all inconveniences of Conventional Neural Networks such as \textit{the lack of interpretability} of the learned parameters, \textit{the resilience to covariance} and \textit{the absence of clear-cut guiding principles} for architecture design.
\par
In \cite{TSnaviga} a general scheme was introduced where  the distinctive and defining property of \textbf{CaNNs} is the following: \textit{the necessary non linearity device that replaces the activation functions in the maps between layers is provided by the exponential map from a solvable Lie algebra to its corresponding solvable Lie group}. This obviously requires that each $i$-th layer of the net should be at the same time a  Riemannian space $(\mathcal{M}_i,g_i)$ endowed with its own metric $g_i$, in order to have geodesics whose lengths provide possible distances between $\mathcal{M}_i$ points,  and a Lie group $\mathcal{S}_i$, necessarily defining its own Lie algebra $\boldsymbol{\mathfrak{s}}_i$. A priori this seems a tight and improbable constraint with very few solutions, yet the combination of four relevant facts that we mention just below, opens a road leading to a rich and well organized landscape of
infinite solutions.
\begin{enumerate}
  \item First we have the priority number one request of Machine Learning, namely that $(\mathcal{M}_i,g_i)$ should admit a unique geodesic joining  any pair of non coincident points so as to have a unique notion of distance between points. In force of Hopf-Rinow theorem \cite{opfarinow} this implies that the layers $(\mathcal{M}_i,g_i)$ are necessarily \textbf{Cartan Hadamard manifolds}, \textit{i.e.} connected Riemannian manifolds with \textit{everywhere non positive sectional curvature} and also leads to \textbf{Cartan Hadamard theorem} stating that any such manifold of dimension $n$ is diffeomorphic to $\mathbb{R}^n$ (see \cite{balmano} for a review).
  \item We have next the second request equally essential for Machine Learning: the  $(\mathcal{M}_i,g_i)$  manifolds should be homogeneous, in order for relative distances to be the only relevant ones. This combines well with the notion of \textbf{Alekseveeskyan normal Riemannian spaces}, namely Cartan Hadamard manifolds where $\mathcal{M}_i=\mathcal{S}_i$ is a \textbf{solvable Lie Group} whose Lie algebra, named  $Solv_i$, is consequently solvable. Furthermore every possible Riemannian metric on $\mathcal{S}_i$ is determined by a \textbf{quadratic non degenerate, positive definite form} $\langle \, , \, \rangle $ on $Solv_i$ that is uniquely transported from the identity element $\boldsymbol{e}\in\mathcal{S}_i$ to any other group element $s\in \mathcal{S}_i$ by group translation \cite{Alekseevsky1975,Cortes,Alekseevsky:2003vw}.
  \item Thirdly, one has the \textbf{systematic metric equivalence} of all \textbf{non-compact symmetric spaces
      $\mathrm{U/H}$}, where  $\mathrm{U}$ is a non-compact semi-simple Lie group and $\mathrm{H}\subset \mathrm{U}$ its maximal compact subgroup, with a unique solvable subgroup $\mathcal{S}_{\mathrm{U/H}}\subset \mathrm{U}$ with the same dimension as the dimension of the coset: $\text{dim}\mathcal{S}_{\mathrm{U/H}} = \text{dim}\mathrm{U/H}$. This metric equivalence is amply reviewed in the foundational paper \cite{pgtstheory} and it is the basis of the \textbf{solvable parameterization}, developed in the years 1997-2007 \cite{Fre:1996js,Andrianopoli:1996bq,Andrianopoli:1997wi,Trigiante:1997zba,pancetta2}, of all the symmetric space geometries of Supergravity Theories.
      All that was also systematically reviewed, in \cite{pgtstheory}, in the perspective of what became the CaNNs setup.
 \item Furthermore there is the systematic classification of all \textbf{Homogeneous Special Geometries} developed through the 1980.s to the early 1990.s within the community of Supergravity Theorists \cite{productproof,SKGaggio3,SKGaggio2,SKGaggio1,specHomgeoA2,specHomgeoA1,vandersuppa,Ferrara:1989py,deWit:1995tf,Lauria:2020rhc}
     (see also the book \cite{Freedman:2012zz}). In such classification we have both non-compact symmetric spaces, \textit{metric equivalent} to a suitable solvable Lie group, and homogeneous manifolds that are not symmetric and, as such have only a solvable Lie group representation. Among the latter we have the L$(-1,P)$ spaces that are the main focus of the present paper.
\item Last but not least, as a relevant component of the Supergravity Legacy to CaNNs, there is the organization of non-compact symmetric spaces $\mathrm{U/H}$ into \textbf{Tits Satake universality classes} and the discovery of the concept of \textbf{Paint Group} $\mathrm{G_{Paint}}$, as outer automorphism group, of the equivalent solvable Lie group $\mathcal{S}_{\mathrm{U/H}}$ \cite{noipainted,titsusataku} (see also the book \cite{advancio} and \cite{pgtstheory} for detailed reviews).
\end{enumerate}
On the basis of all what we recalled above, in \cite{TSnaviga} the conclusion was reached that the most appropriate mathematical modelling of Neural Network layers is provided by non-compact symmetric spaces $\mathrm{U/H}$  and, in particular, that a convenient, yet not exclusive, architecture setup is provided by layers all belonging to the same Tits Satake Universality class. This led to the conception of $r=1,2,..$ CaNNs, labeling the Network with the number of non-compact Cartan generators that, in a Tits Satake Universality class, is common to all members of the class.  A first  explicit realization of an $r=1$ CaNN was provided in \cite{naviga}, whose only  value was the demonstration that, without activation functions and already in the simplest minded architecture design, the \textit{exponential map non linearity} reaches, on standard benchmarks, an accuracy level  that is comparable to that of conventional neural networks. Actually the construction of \cite{naviga} did not attempt at any use of the geometric features of Cartan Neural Networks (Paint Group properties of data representation on the layers, for instance) that, as advocated above and in all previous publications relative to this project, are the main bonus on the road to interpretability of learned parameters. The issue was postponed to $r>1$ CaNNs whose explicit construction had to be done and it is now on its way with a different set of authors \cite{r2paperone} with respect to those of \cite{TSnaviga,naviga}.
\par
In the original perspective of summer 2025, after the completion of papers \cite{TSnaviga,tassellandum}, $r=2$ CaNNs were just the next case after $r=1$, yet the conspicuous developments on Geometric Thermodynamics\cite{gianno1,gianno2,lychaginlecture,ludaed,ludaed2,Lychagin_2020,Kushner_2020},
\cite{Ruppeiner_2010,Ruppeiner_2012,Ruppeiner_2012b,Ruppeiner_2013,ruppoRdiag,Ruppeiner_2020}
and Gibbs distributions \`{a} la Souriau \cite{souriaub1,souriaub2,barbarpapad4,
marlentropia,caldobarbaresco,barbaresco2,barbaresco3,marlegibbs,barbarpapad1,barbarpapad2,barbarpapad3,nebbo}
that were obtained in \cite{geotermico,secondtemperature,terzatemperatura} changed that perspective completely:  $r=2$ CaNNs, rather than a next case, seem now a fixed terminal point. Indeed if we want to combine the vision of \textit{supervised Neural Networks} with the vision of \textit{unsupervised ones with reinforcement}, which requires natural Gibbs-like probability distributions defined over the layers and compatible with their geometric structure, then the latter had better to be K\"ahler manifolds, as discussed at length in \cite{geotermico,secondtemperature,terzatemperatura}.  If moreover we stick to the convenient architectural setup where the layers belong all to the same Tits Satake tower, then the only  Tits Satake  Universality class of K\"ahler symmetric spaces is the Calabi-Vesentini one, namely\footnote{Writing this paper we met a problem of  notations that are conflicting with previous literature in two opposite directions. On one side in the older literature on the classification of homogeneous Special K\"ahler manifolds, which is an essential ingredient of our present topic, the capital letter $P$ was used for what turns out to be the number of paints and the letter $q$ was used for the first indicator in the standard pair of classifying integers $(q,P)$, the classification of the manifolds being based on the structure of Clifford algebras. On the other hand, in all the recent papers of this  project devoted to Calabi-Vesentini manifolds and to their thermodynamics and thermodynamic geometry \cite{geotermico,secondtemperature,terzatemperatura}, the letter $q$, unfortunately, was systematically utilized for the number of paints, while the capital letter $P$, both in roman and in calligraphic form $\mathcal{P}$, was systematically used for the moment maps that  are also going to be  important actors in the expected applications of the present paper to CaNNs. Henceforth in order to avoid conflicts
in both directions, we decided to name $p$ the number of  paints alerting the reader that this $p$ is the $q$ of papers \cite{geotermico,secondtemperature,terzatemperatura}. }:
\begin{equation}\label{Cvmanigoldi}
  \mathcal{M}_{CV}^{[2,p]}\,  \equiv \, \frac{\mathrm{SO}(2,2+p)}{\mathrm{SO}(2)\times \mathrm{SO}(2+p)}
\end{equation}
each of whose members, at all time,  must be regarded as one of the two cofactors constituting an item in following infinite series of Special K\"ahler manifolds:
\begin{equation}\label{speckal}
 \mathcal{SK}_{3+p} \, \equiv \, \frac{\mathrm{SL(2,\mathbb{R})}}{\mathrm{SO(2)}}\times
 \frac{\mathrm{SO}(2,2+p)}{\mathrm{SO}(2) \times \mathrm{SO}(2+p)}
\end{equation}
Actually, one of us,  following the previous results obtained in 1985 with Eugene Cremmer on Special K\"ahler Symmetric Spaces \cite{productproof} proved in 1989, together with Sergio Ferrara \cite{Ferrara:1989py}, that the only Special K\"ahler manifolds that are direct products of two factors are the
$\mathcal{SK}_{3+p}$ defined in eq.(\ref{speckal}).
\par
From the point of view of metric equivalence, crucial for the CaNNs, the CV manifolds (\ref{Cvmanigoldi}) are the following solvable Lie groups
\begin{equation}\label{solsub}
  \mathcal{S}_{[2,2+p]} \, \subset \, \mathrm{SO}(2,2+p)
\end{equation}
whose dimension is equal to that of the symmetric space $\mathcal{M}^{[2,p]}_{CV}$ and whose solvable Lie algebra $Solv_{[2,2+p]}$,  is best described by
the corresponding Maurer Cartan equations. Indeed by introducing  a Paint index  running in the fundamental vector representation of the Paint Group:
\begin{equation}\label{gruppopittura}
  \mathrm{G_{Paint}} \, = \, \mathrm{SO(p)}
\end{equation}
 the MC equations have a simple, general, Paint-invariant form, holding true for the entire TS universality class,  that is the following one: ($\ir=1,\ldots ,p$)
\begin{equation}\label{maurocartus}
Solv_{[2,2+p]} \Leftrightarrow\left\{\begin{array}{lcl}
 \mathrm{d}e^1 & = & 0 \\
 \mathrm{d}e^2 & = & 0 \\
 \mathrm{d}e^3+  e^{1}\wedge e^{3}- e^{2}\wedge e^{3} & = & 0 \\
 \mathrm{d}e^4+  e^{1}\wedge e^{4}+ e^{2}\wedge e^{4}&=&\frac{1}{\sqrt{2}}
 \sum^{p}_{\ir=1} \, e^{5,\ir}\wedge e^{6,\ir} \\
 \mathrm{d}e^{5,\ir}+ e^{1}\wedge e^{5,\ir}+\frac{1}{\sqrt{2}} e^{3}\wedge e^{6,\ir} & = & 0 \\
 \mathrm{d}e^{6,\ir}+e^{2}\wedge e^{6,\ir} & = & 0 \\
\end{array}\right.
\end{equation}
The Alekseevskyan nature of the CV manifolds is made manifest by mentioning the quadratic form on $Solv_{[2,2+p]}$ that defines the symmetric space metric possessing all the isometries of
the $\mathrm{U}$-group $\mathrm{SO}(2,2+p)$. Naming $T_B$ the basis of $Solv_{[2,2+p]}$ dual to the 1-forms $e^A$, namely such that $e^A[T_B]\, = \, \delta^A_B$,  the quadratic form is:
\begin{eqnarray}\label{kappusCV}
  \langle T_A \, , \, T_B\rangle & \equiv & \kappa_{AB}^{CV}\nonumber\\
  \kappa_{AB}^{CV} & = & \text{diag}\left(2,\,2,\,\underbrace{\ft 12,\,\ft 12 ,\dots,\,\ft 12}_{\text{$2+2p$ times}}\right)
  \end{eqnarray}
  Indeed if the  1-forms $e^A$, satisfying the MC equations (\ref{maurocartus}) are identified with left-invariant (or right-invariant) 1-forms on the group manifold $\mathcal{S}_{[2,2+p]}$, parameterized by whatever set of coordinates, then the $\mathrm{SO}(2,2+p)$-isometric line-element can be written as:
  \begin{equation}\label{fracastoro}
   ds^2_{CV} \, = \, \kappa_{AB}^{CV} \, e^A \times e^B
  \end{equation}
 See in particular section 5 of \cite{secondtemperature} for the discussion of the MC equations (\ref{maurocartus}) and the quadratic form (\ref{kappusCV}-\ref{fracastoro}).
\subsection{The result of this paper}
The result that we prove in this paper is that the Special K\"ahler manifolds named L$(-1,P)$ in the classification of all homogenoeus Special Geometries \cite{deWit:1995tf,productproof,SKGaggio3,SKGaggio2,SKGaggio1,specHomgeoA2,specHomgeoA1}
that are homogeneous but not symmetric and correspond to a precise cubic tensor $d_{\ial\jbe\kga}$ (see section \ref{Lm1Pspazio}) for:
\begin{equation}\label{Pqiddo}
  P \, = \, p
\end{equation}
actually correspond to a different K\"ahler metric on the same solvable groups $\mathcal{S}_{[2,2+p]}$ metric equivalent to the CV manifolds (\ref{Cvmanigoldi}). Indeed, parameterizing the left-invariant $1$-forms $e^A$ satisfying the MC eq.s(\ref{maurocartus}) with a set of solvable coordinates
\begin{equation}\label{wcoordi}
  \boldsymbol{w} \, \equiv \, \left\{w^1,\, w^2,\, w^3,\, w^4,\,
   w^{5,1},\, w^{5,2},\,\dots,\,w^{5,p},\,w^{6,1},\, w^{6,2},\,\dots,\,w^{6,p}\right\}
\end{equation}
that will become the coordinates of both the Calabi Vesentini space $\mathcal{M}_{CV}^{[2,p]}$ and of the L$(-1,p)$ homogeneous K\"ahler manifold, while the K\"ahler metric of the former is given by eq.(\ref{fracastoro}), the K\"ahler metric of the latter is given by:
\begin{equation}\label{lm1dsq}
  ds^2_{L[-1,p]} \, = \, \kappa^{L[-1,p]}_{AB} \, e^A \times e^B
\end{equation}
where:
\begin{equation}\label{capponelesso}
  \kappa^{L[-1,q]}_{AB} \, = \, \left(
  \begin{array}{c|c|c}
  \begin{array}{cccc}
 \frac{3}{2} & -\frac{1}{2} & 0 & 0 \\
 -\frac{1}{2} & \frac{3}{2} & 0 & 0 \\
 0 & 0 & 1 & 0 \\
 0 & 0 & 0 & 2 \\
\end{array}& \boldsymbol{0}_{p\times p} & \boldsymbol{0}_{p\times p} \\
\hline
\boldsymbol{0}_{p\times p} & \boldsymbol{1}_{p\times p} &\boldsymbol{0}_{p\times p} \\
\hline
\boldsymbol{0}_{p\times p} & \boldsymbol{0}_{p\times p} &
\frac{1}{2}\boldsymbol{1}_{p\times p} \\
\end{array}
\right)
\end{equation}
As we are going to illustrate below the metric determined by the quadratic form $\kappa^{L[-1,p]}_{AB}$
is invariant with respect to a smaller compact group:
\begin{equation}\label{HLcomp}
  \mathrm{H}^{L[-1,p]}_c \, = \,\mathrm{ U(1)}_L \times \mathrm{SO}(p)
\end{equation}
with respect to
\begin{equation}\label{HCVcomp}
  \mathrm{H}^{CV}_c \, = \,\mathrm{ U(1)}_{CV} \times \mathrm{SO}(2+p)
\end{equation}
\subsection{A subtle point}
\label{sottilettakraft}
As we are going to show below, the compact isometry group of the special K\"ahler manifold L$(-1,p)$ is not a subgroup of the compact isometry group of Calabi-Vesentini manifolds:
\begin{equation}\label{sottiletta}
  \mathrm{H}^{L[-1,p]}_c \, \nsubseteq \, \mathrm{H}^{CV}_c
\end{equation}
because the compact generator $\mathrm{ U(1)}_L$ is not contained in $\mathrm{H}^{CV}_c$. Indeed it acts on
the $1$-forms $e^A$ on which also $\mathrm{H}^{CV}_c$ does act, yet in a way that cannot be reproduced by any element of   $\mathrm{H}^{CV}_c$. One way  to construct $\mathrm{ U(1)}_L$ is described in section \ref{kallergenerator} and it is based on the representation of the solvable Lie algebra $Solv_{[2,2+p]}$ as a subalgebra of $\sym(6+2p,\mathbb{R})$, namely the appropriate symplectic algebra of Special K\"ahler Geometry in the case of L$(-1,p)$ manifolds. Indeed the complete relation replacing eq.(\ref{sottiletta}) is the following one:
\begin{equation}\label{kraftspread}
  \mathbb{H}^{L[-1,p]}_c \, \nsubseteq \, \mathbb{H}^{CV}_c \quad ; \quad \mathbb{H}^{L[-1,p]}_c\bigcap\mathbb{H}^{CV}_c \, = \, \so(p) \, \equiv \, \mathbb{G}_{\mathrm{Paint}}
\end{equation}
that has been written in Lie Algebra rather than in Lie Group notation.
    \subsubsection{The full isometry group \texorpdfstring{$\mathrm{Iso}_{\mathrm{L}(-1,p)}$}{Iso L(-1,p)}}
\label{ciufolino}
There is another way to construct $\mathrm{ U(1)}_L$ which is much more handy and illuminating. Indeed we can realize the entire isometry group $\mathrm{Iso}_{\mathrm{L}(-1,p)}$
of the Special K\"ahler manifold L$(-1,p)$ as a subgroup of
the simple non-compact group $\mathrm{SO}(2,2+p)$ containing all the isometries of the CV symmetric space (\ref{Cvmanigoldi}). \par The abstract structure of the Lie algebra of $\mathrm{Iso}_{\mathrm{L}(-1,p)}$ was determined, together with the symmetry structure of the other special manifolds, in \cite{vandersuppa,deWit:1995tf} (see also \cite{Lauria:2020rhc} and the book \cite{Freedman:2012zz}) briefly summarized in appendix \ref{classmates}, and it is described in terms of Maurer Cartan equations in section \ref{kallergenerator}, see in particular eq.s (\ref{gutenabend}-\ref{maurocartus2}). The full isometry algebra contains in addition to the solvable group generators, $T_A$ and the Paint Group generators $\mathrm{J_{Paint}}^i$ just one more generator $L_-$ that, together with $L_+\equiv T_3$ and $L_0\equiv \ft 12 (T_1 - T_2)$, closes a standard $\mathrm{SL(2,\mathbb{R})}$ as in eq.(\ref{robinia1}). The question is to find an explicit matrix representation of such an algebra and derive the action of each generator on the solvable coordinates that are one-by-one paired  with the generators $T_A$. Such solvable coordinates can be regarded at the same time as the coordinates labeling the points of the symmetric space (\ref{Cvmanigoldi}) or of the  Special K\"ahler $\mathrm{L}(-1,p)$, since as differentiable manifolds and as solvable Lie groups the two spaces are the same. In section
\ref{kallergenerator} we construct the missing $L_-$ as a
$\sym(6+2p)$ symplectic matrix and we show that the $\mathrm{U_L(1)}$ generator $\mathrm{H_L} \equiv L_+ - L_-$ is compact, since its eigenvalues are all either zero or purely imaginary, yet $\mathrm{H_L}$ does not belong to maximal compact subalgebra $\uu(3+p) \subset \sym(6+2p)$.
In section \ref{pseudortho} we determine the extra generator $L_-$ inside the Lie algebra $\so(2,2+p)$ and once again we find that $\mathrm{H_L} \equiv L_+ - L_-$ is compact, yet it does not belong to the maximal compact subalgebra $\mathbb{H}^{CV}_c\, =\, \mathrm{SO(2)\times SO(2+p)}$. Indeed referring to the orthogonal decomposition of the Lie algebra:
\begin{equation}\label{fricandelli}
  \so(2,2+p) \, = \, \mathbb{H}^{CV}_c \oplus \mathbb{K}^{CV}
\end{equation}
we find that the compact operator $\mathrm{H_L}$ is a linear combination of a $\mathbb{K}^{CV}$-element with a suitable $\mathbb{H}^{CV}_c$ element.
\par
Furthermore, as we are going to  demonstrate explicitly, the K\"ahler $2$-form of the L$(-1,p)$ spaces has constant components in the basis of the solvable Lie group $1$-forms $e^A$, namely we find:
\begin{eqnarray}\label{K2formLm1p}
  \mathbb{K} & = & \mathcal{K}_{AB} \, e^A \wedge e^B
\end{eqnarray}
where
\begin{eqnarray}\label{carbonero}
  \mathcal{K}_{AB} & = & \left(
\begin{array}{cccc||ccc|ccc}
 0 & 0 & -1 & 1 & 0 & 0\cdots 0 & 0 & 0 & 0\cdots 0& 0 \\
 0 & 0 & 1 & 1 & 0 & 0\cdots 0 & 0 & 0 & 0\cdots 0 & 0 \\
 1 & -1 & 0 & 0 & 0 & 0\cdots 0 & 0 & 0 & 0\cdots 0& 0 \\
 -1 & -1 & 0 & 0 & 0 & 0\cdots0 & 0 & 0 & 0\cdots 0 & 0 \\
 \hline
 \hline
 0 & 0 & 0 & 0 & 0 & 0\cdots 0 & 0 & -\frac{1}{\sqrt{2}} & 0\cdots 0 & 0 \\
 \begin{array}{c}
 0\\
 \vdots\\
 0\\
 \end{array} & \begin{array}{c}
 0\\
 \vdots\\
 0\\
 \end{array} & \begin{array}{c}
 0\\
 \vdots\\
 0\\
 \end{array} & \begin{array}{c}
 0\\
 \vdots\\
 0\\
 \end{array} & \begin{array}{c}
 0\\
 \vdots\\
 0\\
 \end{array} & \begin{array}{ccc}
  \ddots& & \\
  &\,0\,& \\
  & & \ddots\\
  \end{array} & \begin{array}{c}
 0\\
 \vdots\\
 0\\
 \end{array} & \begin{array}{c}
 0\\
 \vdots\\
 0\\
 \end{array} &\begin{array}{ccc}
  \ddots& & \\
  &-\frac{1}{\sqrt{2}}& \\
  & & \ddots\\
  \end{array} & \begin{array}{c}
 0\\
 \vdots\\
 0\\
 \end{array} \\
 0 & 0 & 0 & 0 & 0 & 0\cdots 0 & 0 & 0 & 0\cdots 0 & -\frac{1}{\sqrt{2}} \\
 \hline
 0 & 0 & 0 & 0 & \frac{1}{\sqrt{2}} & 0\cdots 0 & 0 & 0 & 0\cdots 0 & 0 \\
 \begin{array}{c}
 0\\
 \vdots\\
 0\\
 \end{array} & \begin{array}{c}
 0\\
 \vdots\\
 0\\
 \end{array} & \begin{array}{c}
 0\\
 \vdots\\
 0\\
 \end{array} & \begin{array}{c}
 0\\
 \vdots\\
 0\\
 \end{array} & \begin{array}{c}
 0\\
 \vdots\\
 0\\
 \end{array} & \begin{array}{ccc}
  \ddots& & \\
  &\frac{1}{\sqrt{2}}& \\
  & & \ddots\\
  \end{array} & \begin{array}{c}
 0\\
 \vdots\\
 0\\
 \end{array} & \begin{array}{c}
 0\\
 \vdots\\
 0\\ \end{array} & \begin{array}{ccc}
  \ddots& & \\
  &\,0\,& \\
  & & \ddots\\
  \end{array} & \begin{array}{c}
 0\\
 \vdots\\
 0\\ \end{array} \\
 0 & 0 & 0 & 0 & 0 & 0\cdots 0 & \frac{1}{\sqrt{2}} & 0 & 0\cdots 0 & 0 \\
\end{array}
\right)
\end{eqnarray}
yet the matrix $\mathcal{K}_{AB}$ in eq.(\ref{carbonero}) cannot be interpreted as the adjoint action on the Maurer Cartan of $1$-forms of any one of the  generators of the compact isometry subgroup $\mathrm{H}^{L[-1,p]}_c$ mentioned in eq.(\ref{HLcomp}). From eq.(\ref{carbonero}) we can retrieve a complex structure well defined on the vielbein of the L$(-1,p)$ space, yet this complex structure is not related with any $\mathrm{U(1)}$ subgroup of the compact isometry group. This is probably the most important structural difference between the symmetric Special K\"ahler Manifolds and those that are only homogeneous as L$(-1,p)$. The general formulae for the isometry moment maps utilized in \cite{geotermico,secondtemperature,terzatemperatura}
cease to exist and the construction of Gibbs distributions becomes an open problem. Yet as we summarize in the next subsection, the identification of the two manifolds L$(-1,p)$ and
$\mathcal{M}^{[2,p]}_{CV}$, apart from the metric and the associated complex structure, has important consequences for the architecture of $r=2$ neural networks.
\subsection{Consequences for $r=2$ CaNNs}
\label{conseguenze}
Indeed the consequences for
$r=2$ CaNNs of the identification encoded in eq.s(\ref{lm1dsq}-\ref{capponelesso}) are far reaching. Recalling the general structure of the map between two adjacent layers as described in \cite{TSnaviga} (see eq.s (3.58-3.59)) when we start  the map from layer $i$-th to layer $(i+1)$-th we first reduce an element of the symmetric space $\mathrm{U}_i/\mathrm{H}_i$ to its solvable group $\mathcal{S}_i$ correspondent.
See the scheme here below.
In the case of CV manifolds a convenient rule for this step is the  Cholewsky Crout algorithm that extracts a triangular matrix $\mathbb{L}$ from a symmetric one $M$ in such a way that $M=\mathbb{L}\mathbb{L}^T$. Then with the \textit{group logarithmic map} we reduce this group-element  to an element  $X$ of the solvable Lie algebra $Solv_i$.  By means of the linear embedding/projection map $\mathcal{W}$ (whose parameters are the analogues of the weights)  $X$ is turned into an element $Y$ of the next solvable Lie algebra  $Solv_{i+1}$. At this point we start the reverse path performing on $Y$ the exponential map to the solvable group $\mathcal{S}_{i+1}$. Here because of the identification provided by eq.s(\ref{lm1dsq}-\ref{capponelesso}) we are now confronted with a bifurcation. Indeed we can either map $\mathbb{L}(Y) \, \equiv \, \Sigma[Y]$ to a corresponding element\footnote{We remind the reader that the exponential map from solvable Lie algebra to solvable group can be effected in several different ways that amount to different, obviously equivalent ways of defining the solvable coordinates. For this reason we do not write $\exp[Y]$ rather we write $\Sigma[Y]$ leaving open the choice of a convenient rule for the exponential map.} of the symmetric space $\mathrm{U}_{i+1}/\mathrm{H}_{i+1}$ or to an element of the homogeneous space L$(-1,p_{i+1})$.
\begin{center}
  \begin{tabular}{ccccccccc}
     $\mathrm{U}_i/\mathrm{H}_i $& $\rightarrow $ & $\mathcal{S}_i$ &$ \rightarrow $ &$ Solv_i$ & $\rightarrow $ & $Solv_{i+1}$ & $\rightarrow$  & ${\mathcal S}_{i+1}$\\
     M & $\rightarrow $  &$ \mathbb{L}$ & $\stackrel{\log}{\rightarrow }$ & X & $\stackrel{{\cal W}}{\rightarrow }$ & Y &$ \stackrel{\Sigma }{\rightarrow }$ & $\mathbb{L}(Y)$. \\
   \end{tabular}
\end{center}
\par From a pragmatic point of view the alternative is very simple.
\begin{description}
  \item[a)] In the case of the symmetric space interpretation we just have to build the matrix $M(Y) =\mathbb{L}(Y)\,\mathbb{L}^T(Y)$ and conjugate it:
\begin{equation}\label{corbatosko1}
  M(Y)\to  g\, M(Y)\, g^{-1}
\end{equation}
with a generic compact element:
\begin{equation}\label{corbatosko2}
  g\in \frac{\mathrm{H_c}}{\mathrm{G_{paint}}} \, = \, \mathrm{SO(2)} \times \frac{\mathrm{SO(2+p)}}{\mathrm{SO(2)} \times \mathrm{SO(p)}}
\end{equation}
of the relevant Grassmannian, times a rotation of the $\mathrm{SO(2)} \, \simeq \mathrm{U(1)}$ associated with the K\"ahler structure.
The action of the Grassmannian transformation on the solvable coordinates (i.e. the decomposition coefficients of the solvable Lie algebra element $Y=Y^A T_A$) is obtained via the Cholewsky Crout algorithm, namely by setting:
\begin{equation}\label{carciofus}
   g\, M(Y)\, g^{-1} \, = \, M(g[Y]) \, = \, \mathbb{L}(g[Y])\, \mathbb{L}^T(g[Y])
\end{equation}
  \item[b)] In the case of the L$(-1,p)$ interpretation, thanks to the proof (provided in section \ref{pseudortho}) that the entire isometry group $\mathrm{Iso}_{\mathrm{L}(-1,p)}$ has a faithful representation inside the fundamental representation of $\mathrm{SO}(2,2+p)$, we have nothing more to do a priori since, after the exponential map $\Sigma$, the group element $\mathbb{L}(Y)$ is not only equivalent to, rather it is the very same $(4+p)\times (4+p)$ upper triangular matrix that represents the same point in the symmetric space interpretation. The only consequence of the alternative interpretation is that instead of the isometry (\ref{corbatosko1}-\ref{corbatosko2}) we can perform on
      $\mathbb{L}(Y)$ the unique compact transformation $\mathrm{H_L}$ whose explicit action on the solvable coordinates is derived in section \ref{baldacchino} and is displayed in eq.(\ref{berlinoest}).
\end{description}
It appears then evident that the most economic and effective way of utilizing the double interpretation of the same solvable Lie group in the neural network architecture is that of using both interpretation at each change of layer. After we mapped $Solv_i$ into $Solv_{i+1}$ we make first the unique $\mathrm{H_L}$ rotation with an angle $\theta_{i+1}$, then we insert the obtained solvable coordinates $h_{\theta_{i+1}}[Y]$ into the CV algorithm (\ref{carciofus}) and we obtain:
\begin{equation}\label{malombra}
  g_{i+1}\left[h_{\theta_{i+1}}[Y_i]\right] \, = \, Y_{i+1}
\end{equation}
ready for a new map to the next layer.  Schematically:
\begin{center}
  \begin{tabular}{ccccc}
   $  Solv_{i+1}$ &&&&\\
   $Y_i $&$\rightarrow $&$h_{\theta_{i+1}}[Y_i]$& $\stackrel{g}{\rightarrow }$&$g_{i+1}\left[h_{\theta_{i+1}}[Y_i]\right] \, = \, Y_{i+1}$\,.
   \end{tabular}
\end{center}
\par
The net conclusion is that the double interpretation adds to the CaNN architecture a new (bias) parameter (the angle $\theta$) and a new non linear, structurally different, transformation at every layer transition.
\section{The Special K\"ahler Geometry of L$(-1,P=p)$ and $\mathcal{SK}_{3+p}$ spaces}
\label{simplettica}
For our readers that are outside of the Supergravity/String Theory community we recall that in section 2.1 of \cite{pgtstheory} the mathematical definition of Special K\"ahler geometry (of the local type) was reviewed in a detailed way and the essential role was emphasized of the flat symplectic bundle with structural group $\mathrm{Sp}(2n+2,\mathbb{R})$ where $n= \text{dim}_{\mathbb{C}}\, \mathcal{M}_{n}$ is the complex dimension of the K\"ahler Manifold $\mathcal{M}_{n}$ that is special. The complex dimension of the space L$(-1,p)$ that we define in the next section is:
\begin{equation}\label{blatta1}
 n= \text{dim}_{\mathbb{C}}\left[\mathrm{L}(-1,p)\right]\, = \, 2+p
\end{equation}
so that the relevant symplectic structural group for its Special K\"ahler Geometry is $\mathrm{Sp}(6+2p,\mathbb{R})$. On the other hand the Calabi-Vesentini manifold $\mathcal{M}^{[2,p]}_{CV}$, that has the same complex dimension as L$(-1,p)$:
\begin{equation}\label{blatta2}
  \text{dim}_{\mathbb{C}}\left[\mathcal{M}^{[2,p]}_{CV}\right]\, = \, 2+p
\end{equation}
is not Special by itself yet it is a \textit{codimension one} submanifold of the special K\"ahler manifold $\mathcal{SK}_{3+p}$ (see eq.(\ref{speckal})). Hence the relevant symplectic structural group for $\mathcal{M}^{[2,p]}_{CV}$ manifolds as submanifolds of $\mathcal{SK}_{3+p}$ is $\mathrm{Sp}(8+2p,\mathbb{R})$. In two separate subsections of the present section we review symplectic constructions of the two cases in order to prepare the tools necessary to prove the statements about their relation already anticipated in the introduction.
\subsection{Short summary of the holomorphic symplectic section theory}
\label{sommarioholosect}
From section 2.1 of \cite{pgtstheory} we recall the essential points about the symplectic holomorphic section.
Let ${\cal L} {\stackrel{\pi}{\longrightarrow}} {\cal M}$ denote the complex line bundle whose first Chern class equals the cohomology class of
the K\"ahler form $\mathrm{K}$ of an $n$-dimensional Hodge--K\"ahler manifold ${\cal M}$. Let ${\cal SV} \, \longrightarrow \,{\cal M}$ denote a
holomorphic flat vector bundle of rank $2n+2$ with structural group $\mathrm{Sp}(2n+2,\mathbb{R})$. Consider   tensor bundles of the type ${\cal
H}\,=\,{\cal SV} \otimes {\cal L}$. A typical holomorphic section of such a bundle will be denoted by ${\Omega}$ and will have the following
structure:
\begin{equation}\label{ololo}
  \Omega \, = \,\left( \begin{array}{c}
                   X^\Lambda \\
                   F_ \Sigma
                 \end{array}\right)
   \quad \Lambda,\Sigma =0,1,\dots,n
\end{equation}
By definition the transition functions between two local trivializations $U_i \subset {\cal M}$ and $U_j \subset {\cal M}$ of the bundle ${\cal
H}$ have the following form:
\begin{equation}\label{cambogia}
\left( \begin{array}{c}
                   X^\Lambda \\
                   F_ \Sigma
                 \end{array}\right)_i \, = \, e^{f_{ij}} M_{ij} \left( \begin{array}{c}
                   X^\Lambda \\
                   F_ \Sigma
                 \end{array}\right)_j
\end{equation}
where   $f_{ij}$ are holomorphic maps $U_i \cap U_j \, \rightarrow \,\IC $ while $M_{ij}$ is a constant $\mathrm{Sp}(2n+2,\mathbb{R})$ matrix. For
a consistent definition of the bundle the transition functions are obviously subject to the cocycle condition on a triple overlap:
$e^{f_{ij}+f_{jk}+f_{ki}} = 1 $ and $ M_{ij} M_{jk} M_{ki} = 1 $.
\par
Let ${\rm i}\langle\ \vert\ \rangle$ be the compatible hermitian metric on $\cal H$
\begin{equation}
{\rm i}\langle \Omega \, \vert \, \bar \Omega \rangle \, \equiv \,
- {\rm i} \Omega^\T \left(
\begin{array}{cc}
0 & \mathbf{1} \\
-\mathbf{1} & 0 \\
\end{array}
\right) \,{\bar \Omega} \label{compati}
\end{equation}
\bd We say that a Hodge--K\"ahler manifold ${\cal M}$ is {\bf special K\"ahler} if there exists a bundle ${\cal H}$ of the type described above
such that for some section $\Omega \, \in \, \Gamma({\cal H},{\cal M})$ the K\"ahler two form is given by ($\ial,\jbe=1,\ldots ,n$)
\begin{equation}
\mathrm{K}= \o{\rm i}{2}
 \partial \bar \partial \, \mbox{\rm log} \, \left ({\rm i}\langle \Omega \,
 \vert \, \bar \Omega
\rangle \right )=\frac{i}{2}\,g_{\ial \jbe^*}\,dz^\ial\wedge {\rm d}\bar{z}^{\jbe^*}  \label{compati1}
\end{equation}
\ed From the point of view of local properties, eq.~(\ref{compati1}) implies that we have an expression for the K\"ahler potential in terms of the
holomorphic section $\Omega$:
\begin{equation}
\mathcal{K}\,  = \,  -\mbox{log}\left ({\rm i}\langle \Omega \,
 \vert \, \bar \Omega
\rangle \right )\, =\, -\mbox{log}\left [ {\rm i} \left ({\bar X}^\Lambda F_\Lambda - {\bar F}_\Sigma
X^\Sigma \right ) \right ] \label{specpot}
\end{equation}
In the above formulae (\ref{ololo}-\ref{specpot}) the upper part $X^\Lambda(z)$ and the lower part $F_{\Sigma}(z)$ of the symplectic section are both viewed as holomorphic functions of the complex coordinates $z^\ial$. An important set of coordinates is provided by the \textbf{special coordinates}  that are available when there exists a \textbf{holomorphic homogeneous prepotential} ${F}(\mathbf{X})$ such that
\begin{equation}\label{FLprepot}
F_\Sigma[z] \, = \,  F_\Sigma(X[z]) \, = \, \frac{\partial}{\partial X^\Sigma} F(\mathbf{X})
\end{equation}
Enumerating the section components from $X^0,X^1,\dots,$ to $X^n$ we can use them to define the \textit{special complex coordinates} $z^\ial$ as the following ratios:
\begin{equation}\label{zetaspeciale}
z^\ial\equiv \frac{X^\ial}{X^0}=x^\ial-\rmi\,y^\ial\,
\end{equation}
\subsection{The cubic prepotential for Homogeneous Special K\"ahler Manifolds}
In the papers \cite{deWit:1995tf,productproof,SKGaggio3,SKGaggio2,SKGaggio1,specHomgeoA2}
but in particular in \cite{specHomgeoA1} it was shown that all homogeneous special K\"ahler manifolds, whether symmetric or not, admit a superpotential $F(\mathbf{X})$ that is a cubic polynomial as defined below:
\begin{eqnarray}
    F(\mathbf{X})&\equiv& \frac{1}{6X^0}\,d_{\ial\jbe\kga}\,X^\ial X^\jbe X^\kga \, = \, \left(X^0\right)^2 \, \mathcal{F}(\mathbf{z})\nonumber\\
    \mathcal{F}(\mathbf{z})&\equiv& \frac{1}{6}\,d_{\ial\jbe\kga}\,z^\ial z^\jbe z^\kga
\end{eqnarray}
Fixing $X^0\,=\,1$, that we can always do, since it simply amounts to a K\"ahler transformation on the K\"ahler potential, the symplectic section  (\ref{ololo}) becomes:
\begin{equation}\label{omomo}
  \Omega(\mathbf{z}) \, = \,\left( \begin{array}{c}
                   1 \\
                   z^\ial\\
                   \hline
                   -\mathcal{F}(\mathbf{z})\\
                   \partial_\jbe \mathcal{F}(\mathbf{z})
                 \end{array}\right)
 \quad ; \quad   \ial,\jbe =1,\dots,n
\end{equation}
which inserted into eq.(\ref{specpot}) produces the general K\"ahler potential:
\begin{equation}\label{genKpotyonly}
   \mathcal{K}=-\log\left(\frac{4}{3}\,d_{\ial\jbe\kga}\,y^\ial y^\jbe y^\kga\right)
\end{equation}
Eq.(\ref{genKpotyonly}) reveals two general properties of all homogeneous special geometries:
\begin{description}
  \item[a)] The K\"ahler potential depends only on the imaginary parts of the complex coordinates. This implies a situation amply discussed in \cite{secondtemperature,terzatemperatura} also in relation with geometric thermodynamics, namely there is always a maximal abelian ideal of translational isometries of dimension equal to one half of the real dimension of the manifold.
  \item[b)] All the properties of the metric and its very form are fully determined by the three-index symmetric tensor
      $d_{\ial\jbe\kga}$.
\end{description}
\subsection{The L$(-1,P=p)$ Space}
\label{Lm1Pspazio}
Within the framework outlined above the homogeneous special K\"ahler manifold of type L$(-1,P=p)$, corresponds to the following prepotential  that implicitly defines the components of the symmetric tensor $d_{\ial\jbe\kga}$:
\begin{equation}
    F_{\mathrm{L}(-1,p)}(X) \, = \, \frac{1}{X^0}\left(X^n (X^1)^2-X^1\sum_{\ir=1}^{p} (X^{r+1})^2\right)\,,\qquad \ial,\jbe,\kga=1,\dots, n\,
\end{equation}
where $n=2+p$ is the complex dimension of the manifold as anticipated in eq.(\ref{blatta1}).
The automorphism of the solvable Lie algebra is
\begin{equation}\label{Gpittura}
 \mathrm{ G_{Paint}}\, = \,{\rm SO}(p)
\end{equation}
The non-vanishing entries of the $d_{\ial\jbe\kga}$ tensor are: $d_{11n}=2=-d_{1\ir\ir}$.
Using special coordinates $z^i$ as in eq.(\ref{zetaspeciale}),
the K\"ahler potential has the general form (\ref{genKpotyonly}).
\par
The solvable parametrization is obtained from the one of the real special counterpart $L_R(-1,p)$. Indeed in the series of papers \cite{deWit:1995tf,productproof,SKGaggio3,SKGaggio2,SKGaggio1,specHomgeoA2}
the classification of homogeneous special geometries was elaborated together with the discovery of the maps from Real Special Geometry to Special K\"ahler Geometry, named the $R$-map and from Special K\"ahler Geometry to Quaternionic K\"ahler Geometry named the $c$-map (see detailed reviews in \cite{pgtstheory} and in the books \cite{Lauria:2020rhc,Freedman:2012zz,advancio}). The guiding principle was the dimensional reduction of supergravity theories first from $5$ to $4$-dimensions, then from $4$ to $3$-dimensions.
\par In this conceptual framework the real predecessor of L$(-1,p)$ is the real manifold $L_R(-1,p)$ containing the scalars of the $D=5$ vector multiplets.  The variety $L_R(-1,p)$ is immersed in $\mathbb{R}^n$ as the locus of points, of real coordinates $h^\ial$, satisfying the equation:
\begin{equation}\label{LRdefi}
  \frac{1}{6}\,d_{\ial\jbe\kga}\,h^\ial h^\jbe h^\kga=1\,.
\end{equation}
When going from $L_R(-1,p)$ to L$(-1,p)$, we identify:
\begin{equation}
    y^\ial=e^\psi\,h^\ial\,
\end{equation}
where $\psi$ is the scalar related to the K\"ahler potential $\mathcal{K}$:
\begin{equation}
   \frac{1}{8}\, e^{-\mathcal{K}}=e^{3\psi}\,
\end{equation}
The explicit form of $y^\ial$ in the solvable parametrization is:
\begin{align}\label{ytoLm1solv}
y^1&=e^{\frac{\varphi }{\sqrt{6}}+\psi }\,,\,\,\,y^\ir=\frac{1}{\sqrt{2}} \chi^\ir\,{ e^{\frac{\varphi }{\sqrt{6}}+\psi }}\,,\,\,\,y^n=e^\psi\,\left(\frac{1}{2} e^{\frac{\varphi }{\sqrt{6}}} |\boldsymbol{ \chi}|^2+e^{-\sqrt{\frac{2}{3}} \varphi }\right)\, \quad ; \quad \ir=1,\dots,p
\end{align}
where $|\boldsymbol{ \chi}|^2\equiv\sum_{\ir=1}^{p} (\chi^\ir)^2 $.
The metric
\begin{equation}
{\rmd}s^2=2\,g_{\ial\bar{\jbe}}\,dz^\ial\,\rmd \bar{z}^{\bar{\jbe}}\,\,,\,\qquad
\,g_{\ial\bar{\jbe}}=\partial_\ial\partial_{\bar{\jbe}}\mathcal{K}\,,
\end{equation}
has the following explicit form:
\begin{align}\label{fortedeipietroni}
    {\rmd} s^2_{\mathrm{L}(-1,p)}&=\frac{1}{2}\,{\rmd}\varphi^2+\frac{3}{2}\,{\rmd\psi^2}
    +\frac{1}{2}e^{-\sqrt{\frac{3}{2}}\varphi- 2 \psi}\,(\rmd x^1)^2+e^{\frac{\varphi}{\sqrt{6}}- 2 \psi}\,|\rmd{\bf x}|^2+\frac{1}{2}\,e^{\sqrt{\frac{3}{2}}\varphi}\,|\rmd\boldsymbol{\chi}|^2
    -e^{\frac{\varphi}{\sqrt{6}}- 2 \psi}\,\rmd x^1 \rmd x^n+\nonumber\\
  &+\frac{1}{2}\,e^{2 \sqrt{\frac{2}{3}} \varphi -2 \psi } \left[ \left(\frac{1}{2} \,|\boldsymbol{ \chi}|^2+ e^{-\sqrt{\frac{3}{2}}\varphi }\right)\rmd x^1-\sqrt{2}\, \rmd{\bf x}\cdot \boldsymbol{\chi}+\rmd x^n\right]^2\,,
\end{align}
where we have introduced the vectors ${\bf x}\equiv (x^{r+1})_{r=1,\dots, p}$ and $\boldsymbol{\chi}=(\chi^r)_{r=1,\dots, p}$.
\subsubsection{The solvable Lie algebra inside $\Sp(6+2p,\mathbb{R})$}
\label{solvableinsp6}
Our next task is that of writing a set of generators, all belonging to Lie algebra $\sym(6+2p,\mathbb{R})$ that close a solvable subalgebra $Solv_{\mathrm{L}(-1,p)}$ of dimension equal to the real dimension of $\mathrm{L}(-1,p)$, namely $2\,n$  and such that the metric (\ref{fortedeipietroni}) can be rewritten as a quadratic symmetric form with constant coefficients in the left invariant $1$-forms $V^A$ ($A=1,\dots, 2\,n$) of the corresponding solvable Lie group $\mathcal{S}_{\mathrm{L}(-1,p)}$.
\par
Before doing that and in order to accomplish later the task of constructing the extra compact generator $\mathrm{H}_L$ it is convenient to summarize the structure of generic elements of the
Lie algebra $\sym(6+2p,\mathbb{R})$. This is done in appendix
\ref{simplettando} to which we refer the reader for details.
\paragraph{\sc The solvable Lie algebra $Solv_{\mathrm{L}(-1,p)}\subset
\sym(6+2p,\mathbb{R})$.}
Let us write the generators $t_A$ of the solvable Lie algebra $Solv_{\mathrm{L}(-1,p)}$ as follows:
\begin{equation}
    Solv_{\mathrm{L}(-1,p)}=\text{span}\left[\underbrace{t_A}_{A=1,\dots,2\,n}\right]\, \equiv\, \text{span}\left[\underbrace{h_\phi,\,h_\psi,\,E_{\chi^\ir},\,E_{x^\ial}}_{2\,n}\right]\,,\,\,\,\ir=1,\dots, p\,,\,\,\ial=1,\dots , n=2+p\,
\end{equation}
The gradings are defined with respect to the Cartan generators ${\bf h}=\{h_\phi,\,h_\psi\}$:
\begin{align}\label{lagradisca}
    [{\bf a}\cdot {\bf h},\,E_{\chi^\ir}]&=\,\sqrt{\frac{3}{2}}\,a_1\,E_{\chi^\ir}\,,\nonumber\\
       [{\bf a}\cdot {\bf h},\,E_{x^1}]&=\left(-\sqrt{\frac{2}{3}} a_1-2 a_2\right)\,E_{x^1}\,,\nonumber\\
         [{\bf a}\cdot {\bf h},\,E_{x^{\ir+1}}]&=\left(\frac{a_1}{\sqrt{6}}-2 a_2\right)\,E_{x^{\ir+1}}\,,\nonumber\\
           [{\bf a}\cdot {\bf h},\,E_{x^{n}}]&=\left(2 \sqrt{\frac{2}{3}} a_1-2 a_2\right)\,E_{x^{n}}\,.
\end{align}
In a suitable basis (as in  (\ref{ololo}) with $\Lambda ,\Sigma =0,1,\ir,n$), the generators have the following form as $\sym(6+2p,\mathbb{R})$ matrices:
\begin{align}\label{gennih1}
   h_\phi&=\left(
\begin{array}{cccccccc}
 0 & 0 & 0 & 0 & 0 & 0 & 0 & 0 \\
 0 & \sqrt{\frac{2}{3}} & 0 & 0 & 0 & 0 & 0 & 0 \\
 0 & 0 & -\frac{1}{\sqrt{6}}\,{\bf 1}_{p\times p} & 0 & 0 & 0 & 0 & 0 \\
 0 & 0 & 0 & -2 \sqrt{\frac{2}{3}} & 0 & 0 & 0 & 0 \\
 0 & 0 & 0 & 0 & 0 & 0 & 0 & 0 \\
 0 & 0 & 0 & 0 & 0 & -\sqrt{\frac{2}{3}} & 0 & 0 \\
 0 & 0 & 0 & 0 & 0 & 0 & \frac{1}{\sqrt{6}} \,{\bf 1}_{p\times p}& 0 \\
 0 & 0 & 0 & 0 & 0 & 0 & 0 & 2 \sqrt{\frac{2}{3}} \\
\end{array}
\right)\,
\end{align}
\begin{align}\label{gennih2}
 h_\psi&=\left(
\begin{array}{cccc}
 -3 & 0 & 0 & 0 \\
 0 & -{\bf 1}_{n\times n} & 0 & 0 \\
 0 & 0 & 3 & 0 \\
 0 & 0 & 0 &{\bf 1}_{n\times n} \\
\end{array}
\right)
\end{align}
\begin{align}\label{gennichi}
\chi^r\,E_{\chi_r}&=\left(\begin{matrix}A & {\bf 0}\cr {\bf 0} & -A^t \end{matrix}\right)\,,\,\,A=\left(\begin{matrix}0 & 0 & {\bf 0}_p &0\cr 0 & 0 & \sqrt{2}\boldsymbol{\chi} &0\cr {\bf 0}_p & {\bf 0}_p & {\bf 0}_{p\times p} & 2\sqrt{2}\,\boldsymbol{\chi}^t\cr 0 & 0 & {\bf 0}_p &0\end{matrix}\right)
\end{align}
\begin{align}\label{gennix}
x^\ial\,E_{x^\ial}&=\left(\begin{matrix}B & C\cr {\bf 0} & -B^t \end{matrix}\right)\,,\,\,\,B=\left(\begin{matrix}0 & 2\boldsymbol{\xi}\cr {\bf 0}_{n}& {\bf 0}_{n\times n}\end{matrix}\right)\,,\,\,\,C=\left(\begin{matrix}0 & {\bf 0}_n\cr {\bf 0}_n & d_{\ial\jbe\kga}\,\xi^\kga\end{matrix}\right)\,,
\end{align}
where we have introduced the following normalization of the vector $\boldsymbol{\xi}$
\begin{equation}\label{felicetti}
  \boldsymbol{\xi}=(\xi^\ial)\equiv (x^1,\,-2 {\bf x},\,4\,x^n)
\end{equation}
which amounts to a normalization of the solvable Lie algebra generators introduced above with respect to the real parts of the original complex coordinates in (\ref{zetaspeciale}).
\subsubsection{The left invariant $1$-forms of the solvable Lie group $\mathcal{S}_{\mathrm{L}(-1,p)}$}
Next, we reconstruct the metric (\ref{fortedeipietroni}) from the solvable group manifold $\mathcal{S}_{\mathrm{L}(-1,p)}\equiv\exp[Solv_{\mathrm{L}(-1,p)}]$, defining first our choice of the exponential map
\begin{eqnarray}\label{sigmamapgen}
  \Sigma & : & Solv_{\mathrm{L}(-1,p)} \, \longrightarrow \, \mathcal{S}_{\mathrm{L}(-1,p)}
\end{eqnarray}
explicit form,
which just amounts to a choice of the solvable coordinates. We set:
\begin{eqnarray}\label{sigmamapours}
\forall  X &=&
\underbrace{\varphi\,h_{\varphi}+\psi\,h_\psi}_{\text{Cartan}}
\,+\,
\underbrace{\chi^r\,E_{\chi_r}}_{ \text{$\chi$-nilpotent}}
+\underbrace{x^\ial E_{x^\ial}}_{\text{max. ab. ideal}} \, \in Solv_{\mathrm{L}(-1,p)}
\nonumber\\
\Sigma\left[X\right] &=&\mathbb{L}_d\left(\varphi,\psi,\mathbf{x},\boldsymbol{\chi}\right) \, \equiv \,\underbrace{\exp\left[x^\ial E_{x^\ial}\right]}_{\mathbb{L}_{\mathbf{x}}}\cdot \underbrace{\exp\left[\chi^\ir\,E_{\chi^\ir}\right]}_{\mathbb{L}_{\boldsymbol{\chi}}}
\cdot\underbrace{\exp\left[-\ft 12\,\varphi h_{\varphi}-\ft 12 \,\psi\,h_\psi\right]}_{\mathbb{L}_{\boldsymbol{\varphi}\boldsymbol{\psi}}}
\end{eqnarray}
In order to illustrate the explicit structure of the solvable group $\mathcal{S}_{\mathrm{L}(-1,p)}$  in its symplectic representation
we choose $p=3$ which yields $n=5$ and we display the three
$12 \times 12$ matrices $\mathbb{L}_{\mathbf{x}}$, $\mathbb{L}_{\boldsymbol{\chi}}$, $\mathbb{L}_{\boldsymbol{\varphi}\boldsymbol{\psi}}$ composing the generic group element $\mathbb{L}_d$ as defined in eq.(\ref{sigmamapours})
{\scriptsize
\begin{eqnarray}\label{Lx}
    &\mathbb{L}_{\mathbf{x}}=& \nonumber\\
    & \left(
\begin{array}{cccccc|cccccc}
 1 & 2 x^1 & -4 x^2 & -4 x^3 & -4 x^4 & 8 x^5 & 16 x^1 \left(\vec{x}^2-x_1 x_5\right) & -8 \left(\vec{x}^2-2 x^1 x^5\right) & 8 x^1 x^2 & 8
   x^1 x^3 & 8 x^1 x^4 & 2 (x^1)^2 \\
 0 & 1 & 0 & 0 & 0 & 0 & 8 \left(\vec{x}^2-2 x^1 x^5\right) & 8 x^5 & 4 x^2 & 4 x^3 & 4 x^4 & 2 x^1 \\
 0 & 0 & 1 & 0 & 0 & 0 & -8 x^1 x^2 & 4 x^2 & -2 x^1 & 0 & 0 & 0 \\
 0 & 0 & 0 & 1 & 0 & 0 & -8 x^1 x^3 & 4 x^3 & 0 & -2 x^1 & 0 & 0 \\
 0 & 0 & 0 & 0 & 1 & 0 & -8 x^1 x^4 & 4 x^4 & 0 & 0 & -2 x^1 & 0 \\
 0 & 0 & 0 & 0 & 0 & 1 & -2 (x^1)^2 & 2 x^1 & 0 & 0 & 0 & 0 \\
 \hline
 0 & 0 & 0 & 0 & 0 & 0 & 1 & 0 & 0 & 0 & 0 & 0 \\
 0 & 0 & 0 & 0 & 0 & 0 & -2 x^1 & 1 & 0 & 0 & 0 & 0 \\
 0 & 0 & 0 & 0 & 0 & 0 & 4 x^2 & 0 & 1 & 0 & 0 & 0 \\
 0 & 0 & 0 & 0 & 0 & 0 & 4 x^3 & 0 & 0 & 1 & 0 & 0 \\
 0 & 0 & 0 & 0 & 0 & 0 & 4 x^4 & 0 & 0 & 0 & 1 & 0 \\
 0 & 0 & 0 & 0 & 0 & 0 & -8 x^5 & 0 & 0 & 0 & 0 & 1 \\
\end{array}
\right)&\nonumber\\
&\mbox{with } \vec{x}^2=(x^2)^2+(x^3)^2+(x^4)^2\,,\nonumber\\
\end{eqnarray}
}
{\scriptsize
\begin{eqnarray}\label{Lchi}
    &\mathbb{L}_{\boldsymbol{\chi}}=& \nonumber\\
    & \left(
\begin{array}{cccccc|cccccc}
 1 & 0 & 0 & 0 & 0 & 0 & 0 & 0 & 0 & 0 & 0 & 0 \\
 0 & 1 & \sqrt{2} \chi^1 & \sqrt{2} \chi ^2 & \sqrt{2} \chi ^3 & 2 \left((\chi ^1)^2+(\chi ^2)^2+(\chi ^3)^2\right) & 0 & 0 & 0 & 0 & 0 & 0 \\
 0 & 0 & 1 & 0 & 0 & 2 \sqrt{2} \chi^1 & 0 & 0 & 0 & 0 & 0 & 0 \\
 0 & 0 & 0 & 1 & 0 &  2 \sqrt{2}\chi ^2 & 0 & 0 & 0 & 0 & 0 & 0 \\
 0 & 0 & 0 & 0 & 1 &  2 \sqrt{2} \chi ^3 & 0 & 0 & 0 & 0 & 0 & 0 \\
 0 & 0 & 0 & 0 & 0 & 1 & 0 & 0 & 0 & 0 & 0 & 0 \\
 \hline
 0 & 0 & 0 & 0 & 0 & 0 & 1 & 0 & 0 & 0 & 0 & 0 \\
 0 & 0 & 0 & 0 & 0 & 0 & 0 & 1 & 0 & 0 & 0 & 0 \\
 0 & 0 & 0 & 0 & 0 & 0 & 0 & -\sqrt{2}\chi ^1 & 1 & 0 & 0 & 0 \\
 0 & 0 & 0 & 0 & 0 & 0 & 0 & -\sqrt{2} \chi ^2 & 0 & 1 & 0 & 0 \\
 0 & 0 & 0 & 0 & 0 & 0 & 0 & -\sqrt{2} \chi ^3 & 0 & 0 & 1 & 0 \\
 0 & 0 & 0 & 0 & 0 & 0 & 0 & 2 \left((\chi ^1)^2+(\chi ^2)^2+(\chi ^3)^2\right) & -2 \sqrt{2}\chi ^1 & -2 \sqrt{2} \chi ^2 & -2 \sqrt{2} \chi ^3 & 1 \\
\end{array}
\right)&\nonumber\\
\end{eqnarray}
}
{\fontsize{3.6}{3.8}
\begin{eqnarray}\label{Lvarphipsi}
&\mathbb{L}_{\boldsymbol{\varphi\psi}}=& \nonumber\\
&\left(\begin{smallmatrix}
  e^{3 \psi /2} & 0 & 0 & 0 & 0 & 0 & 0 & 0 & 0 & 0 & 0 & 0 \\
 0 & e^{\frac{\psi }{2}-\frac{\varphi }{\sqrt{6}}} & 0 & 0 & 0 & 0 & 0 & 0 & 0 & 0 & 0 & 0 \\
 0 & 0 & e^{\frac{1}{12} \left(\sqrt{6} \varphi +6 \psi \right)} & 0 & 0 & 0 & 0 & 0 & 0 & 0 & 0 & 0 \\
 0 & 0 & 0 & e^{\frac{1}{12} \left(\sqrt{6} \varphi +6 \psi \right)} & 0 & 0 & 0 & 0 & 0 & 0 & 0 & 0 \\
 0 & 0 & 0 & 0 & e^{\frac{1}{12} \left(\sqrt{6} \varphi +6 \psi \right)} & 0 & 0 & 0 & 0 & 0 & 0 & 0 \\
 0 & 0 & 0 & 0 & 0 & e^{\sqrt{\frac{2}{3}} \varphi +\frac{\psi }{2}} & 0 & 0 & 0 & 0 & 0 & 0 \\
 0 & 0 & 0 & 0 & 0 & 0 & e^{-3 \psi /2} & 0 & 0 & 0 & 0 & 0 \\
 0 & 0 & 0 & 0 & 0 & 0 & 0 & e^{\frac{\varphi }{\sqrt{6}}-\frac{\psi }{2}} & 0 & 0 & 0 & 0 \\
 0 & 0 & 0 & 0 & 0 & 0 & 0 & 0 & e^{\frac{1}{12} \left(-\sqrt{6} \varphi -6 \psi \right)} & 0 & 0 & 0 \\
 0 & 0 & 0 & 0 & 0 & 0 & 0 & 0 & 0 & e^{\frac{1}{12} \left(-\sqrt{6} \varphi -6 \psi \right)} & 0 & 0 \\
 0 & 0 & 0 & 0 & 0 & 0 & 0 & 0 & 0 & 0 & e^{\frac{1}{12} \left(-\sqrt{6} \varphi -6 \psi \right)} & 0 \\
 0 & 0 & 0 & 0 & 0 & 0 & 0 & 0 & 0 & 0 & 0 & e^{\frac{1}{6} \left(-2 \sqrt{6} \varphi -3 \psi \right)} \\
\end{smallmatrix}\right)& \nonumber\\
\end{eqnarray}
}
Next, we compute the left-invariant 1-forms:
\begin{equation}
  \mathbb{L}_d^{-1}\mathrm{d}\mathbb{L}_d=V^A\,t_A\,.
\end{equation}
Explicitly we obtain:
\begin{alignat}{5}\label{Viforme}
  V^1 &\, = \,&\, - \, &\ft 12 \, \mathrm{d}\varphi &\null&\nonumber\\
   V^2 &\, = \,&\, - \, &\ft 12 \, \mathrm{d}\psi &\null&\nonumber\\
 V^{2+r}&\, = \,&\,  & e^{\frac{1}{2} \sqrt{\frac{3}{2}} \varphi } \mathrm{d}\chi^\ir  &\quad;\quad &\ir=1,\dots, p\nonumber\\
 V^{3+p} &\, = \,&\, \null \, & e^{-\frac{\varphi }{\sqrt{6}}-\psi }\,\mathrm{d}x^1 &\null&\nonumber\\
 V^{3+p+r}&\, = \,&\, & e^{\frac{\varphi }{2 \sqrt{6}}-\psi } \left(-\frac{1}{\sqrt{2}} \,\chi^r \, \mathrm{d}x^1 +\, \mathrm{d}x^\ir\right) &\quad;\quad &
 \nonumber\\
 V^{4+2p} &\, = \,&\, \null \, & e^{\sqrt{\frac{2}{3}} \varphi -\psi } \left(\ft 12\mathrm{d}x^1 \,\sum_{\ir=1}^p (\chi^\ir)^2\,-{\sqrt{2}}\sum_{\ir=1}^p \chi^\ir \,\mathrm{d}x^{1+\ir}
  +\rmd x^{n}\right)&\null&
\end{alignat}
\subsubsection{The quadratic form on $Solv_{\mathrm{L}(-1,p)}$}
Introducing the following symmetric constant tensor on the solvable Lie algebra $Solv_{\mathrm{L}(-1,p)}$:
\begin{equation}\label{kappalone}
\langle\,t_A \, , \, t_B\,\rangle \, = \,  \boldsymbol{\mathfrak{q}}_{AB} \, \equiv \, \text{diag}\left[2,6,\underbrace{\frac{1}{2},\cdots,\frac{1}{2}}_{p \,\text{ entries}},1,\underbrace{1,\cdots,1}_{p \,\text{ entries}},\frac{1}{2}\right]_{AB}
\end{equation}
we easily verify that the metric in eq.(\ref{fortedeipietroni}) is exactly reproduced by the following position:
\begin{equation}\label{colabrodo}
    \rmd s^2_{L(-1,q)}=V^A\times V^B\,\boldsymbol{\mathfrak{q}}_{AB}\,
\end{equation}
\subsubsection{The Maurer Cartan equations satisfied by the $1$-forms $V^A$}
The next fundamental step towards the proof of our statement (\ref{lm1dsq}) consists of the derivation of the Maurer Cartan equations satisfied by the $1$-forms (\ref{Viforme}), which is the same thing as deriving the structure constants in the commutation relations of the generators $t_A$, that we have explicitly constructed as $\sym(6+2p,\mathbb{R})$ symplectic matrices.
\par
Explicitly we find:
\begin{alignat}{5}
\label{mceqlm1q}
  &\mathrm{d}V^1 &\,\,=\,\,& 0 &\null &\null \nonumber\\
  &\mathrm{d}V^2 &\,\,=\,\,& 0 &\null &\null\nonumber\\
 &\mathrm{d}V^{2+\ir}\, + \, \sqrt{\frac{3}{2}} \, V^1\wedge V^{2+\ir} &\,\,=\,\,& 0 &\quad ; \quad & 
 \nonumber\\
&\mathrm{d}V^{3+p}\, - \, \sqrt{\frac{2}{3}} \, V^1\wedge V^{3+p}\, - \, 2 \, V^2\wedge V^{3+p}&\,\,=\,\,& 0 &\null & \null \nonumber\\
&\mathrm{d}V^{3+p+\ir}\, + \, \frac{1}{\sqrt{6}} \, V^1\wedge V^{3+p+\ir}\,- \, 2 \, V^2\wedge V^{3+p+\ir} \, + \, \frac{1}{\sqrt{2}}\,V^{2+\ir} \wedge V^{3+p}&\,\,=\,\,& 0 &\quad ; \quad & 
\nonumber\\
&\mathrm{d}V^{2n}\, + \, 2\,\sqrt{\frac{2}{3}} \, V^1\wedge V^{2n}\, - \, 2 \, V^2\wedge V^{2n}\, + \, \sqrt{2}\, \sum_{\ir=1}^p \,V^{2+\ir} \wedge V^{3+p+r}&\,\,=\,\,& 0 &\null & \null
\end{alignat}
\subsubsection{The isomorphism of $Solv_{\mathrm{L}(-1,p)}$ and $Solv_{[2,2+p]}$ of the CV manifolds}
Eq.s (\ref{mceqlm1q}) identify the solvable Lie algebra $Solv_{\mathrm{L}(-1,p)}$ as isomorphic to the solvable Lie algebra
$Solv_{[2,2+p]}$ underlying Calabi-Vesentini symmetric spaces whose structure is exhibited in eq.(\ref{maurocartus}).
To see the isomorphism explicitly it suffices to perform the following identification between the $1$-forms $V^A$ and the $1$-forms $e^A$:
\begin{alignat}{5}
\label{finocchiona}
  V^1 &\,=\,& \sqrt{\frac{2}{3}} \, e^2 &\null & \null\nonumber\\
  V^2 &\,=\,& \frac{1}{6} \,\left(- 3\, e^1  + e^2\right) &\null & \null \nonumber\\
  V^{2+\ir} &\,=\,& e^{6,\ir} &\quad ; \quad & 
  \nonumber\\
 V^{3+p} &\,=\,& - e^{3} &\null & \null\nonumber\\
  V^{3+p+\ir} &\,=\,& e^{5,\ir} &\quad ; \quad & 
  \nonumber\\
   V^{2n} &\,=\,& 2 e^{4} &\null & \null
\end{alignat}
Upon the substitution (\ref{finocchiona}), eq.s(\ref{mceqlm1q}) transform into equations (\ref{maurocartus}) and eq.(\ref{colabrodo}) is transformed into eq.(\ref{lm1dsq}).
\subsection{The Special K\"ahler Geometry of $\mathcal{SK}_{3+p}$}
Having established the above identification it is now clear that we can rewrite the Special K\"ahler Geometry of L$(-1,p)$ in terms of the same solvable coordinates that we utilize for the Calabi-Vesentini manifolds: in other words we can find the relation between the coordinates $y^i,x^i$ used above and the corodinates $w^1,\ldots,w^4,w^{5,\ir},w^{6,\ir}$ utilized in the CV case. It is convenient for perspective subsequent developments to introduce also the $\mathrm{SL(2,\mathbb{R})/SO(2)}$ cofactor and consider the complete parameterization of the Special K\"ahler manifold
$\mathcal{SK}_{3+p}$.
\subsubsection{Solvable parameterization of the CV manifolds}
\label{sciogliCV}
For the CV manifolds we use the same conventions and coordinate choices used in \cite{secondtemperature} and also in \cite{pgtstheory}\footnote{Since behind the whole series of papers \cite{pgtstheory,TSnaviga,naviga,tassellandum,axialbeltra,geotermico,secondtemperature,terzatemperatura} there is always the same WOLFRAM Mathematica code \textbf{CVPlatform.nb} that has evolved  over the last three years, by addition and inclusion, into a rather monumental software with User's Guide, definitions and examples, based  on tens over tens of routines, subroutines, modules and newly defined functions, we prefer also in the publications to stick to the normalization of Lie Algebra generators, solvable coordinates and so on that are built in the Mathematica code \textbf{CVPlatform.nb}. This guarantees that all the displayed formulae are aligned with those that we produce with the code and can be checked at any moment. Moreover it guarantees that the formulae cited from previous papers of the series are coherent with the new ones introduced in each new paper. Furthermore this is very much relevant for the next steps in our research programme that foresee the explicit construction in numerical codes like Python of CaNNs \cite{r2paperone}.
The exportation from Mathematica to Python of the algorithm building blocks requires that the normalizations be always the same. Names of the indices and of the symbols can change from paper to paper according to the theoretical needs, this is easy to control. On the contrary changes of normalizations are almost impossible in such a large Mathematica Code without introducing mistakes or conflicts that jeopardize the proper functioning of the calculation code. }. We begin by writing the solvable generators in the fundamental representation of
$\mathrm{SO}(2,2+p)$ namely in terms of $(4+2p)\times(4+2p)$ matrices, utilizing as invariant $\eta$-tensor the one of the so
called \textbf{triangular basis} (see \cite{pgtstheory}) in which all solvable elements of the $\so(2,2+p)$ Lie algebra are upper triangular. The matrix $\eta_t$ is defined in eq.(5.25) of \cite{secondtemperature}\footnote{See also section 3 of \cite{pgtstheory} eq.(3.45).} and we recall it here for reader's convenience:
\begin{equation}\label{etatdefi}
  \eta_t \, = \, \left(
\begin{array}{cc|c|cc}
 0 &  0 & 0 & 0 &  1 \\
 0 &  0 & 0 & 1 &  0 \\
 \hline
 \mathbf{0}_{p\times 1} &  \mathbf{0}_{p\times 1} & \mathbf{1}_{p \times p} & \mathbf{0}_{p\times 1} & \mathbf{0}_{p\times 1}  \\
 \hline
  0 &  1 & 0 & 0 &  0 \\
 1 &  0 & 0 & 0 &  0 \\
\end{array}
\right)
\end{equation}
Hence, as stated in eq.(5.26) of \cite{secondtemperature}, the group $\mathrm{SO}(2,2+p)$ is defined as the set of all non singular $(4+2p)\times(4+2p)$ matrices that preserve $\eta_t$:
\begin{equation}\label{invametra4piuq}
  \mathrm{SO}(2,2+p) \, = \, \left\{L \in \mathrm{GL(4+2p,\mathbb{R})} \, \mid \,
  L^T \cdot \eta_t \cdot L \, = \, \eta_t\right\}
\end{equation}
and the solvable subgroup $\mathcal{S}_{[2,2+p]} \subset \mathrm{SO}(2,2+p)$ is the subset in $\mathrm{SO}(2,2+p)$ of all those matrices that are upper triangular. The utilized basis of
generators for the solvable Lie algebra $Solv_{[2,2+p]}$ is described both in the appendices of \cite{secondtemperature} and in the main body of \cite{pgtstheory}.
\par
The choice of  the form of the exponential map
\begin{eqnarray}\label{sigmamapgenCV}
  \Sigma_{CV} & : & Solv_{[2,2+p]} \, \longrightarrow \, \mathcal{S}_{[2,2+p]}
\end{eqnarray}
is encoded in the definition of the solvable coset representative $\mathbb{L}(\mathbf{W})$ which is that presented in eq.(6.3) of \cite{secondtemperature}, once again repeated here for reader's convenience\footnote{We remind the reader that for the already mentioned notation conflict the letter $p$ denotes here what was noted in \cite{secondtemperature} and \cite{pgtstheory} with the letter $q$.}:
\begin{equation}\label{krollus}
  \mathbb{L}\left(\mathbf{W}\right)\, = \, \underbrace{\exp\left[\sum^{q}_{\ir=1}\,w^{5,\ir}\, T_{5,\ir} + w^3\,T_3 + w^4\,  T_4\right]}_{\text{max. ab. id.}} \cdot \underbrace{\exp\left[w^1\,T_1 + w^2 \, T_2\right]}_{\text{non comp. Cartan}}\cdot\underbrace{\exp\left[\sum^{p}_{\ir=1}\,w^{6,\ir}\, T_{6,\ir}\right]}_{\text{compl. ab. subalg.}}
\end{equation}
With such a definition the explicit expression of the Maurer Cartan $1$-forms satisfying the MC equations (\ref{maurocartus})is the following:
\begin{alignat}{4}
\label{econdor}
  e^1 &\,=\, \mathrm{d}w^1 &\null & \null&\nonumber\\
  e^2 &\,=\, \mathrm{d}w^2 &\null & \null &\nonumber\\
  e^{3} &\,=\, e^{w^2-w^1} \mathrm{d}w^3 &\null & \null&\nonumber\\
   e^{4} &\,=\, \frac{1}{4}\, e^{-w^1-w^2} \left(-e^{2 w^2}
   \mathrm{d}w^3 \left(\sum_{\ir=1}^p (w^{6,\ir})^2\right)- 2 \sqrt{2} e^{w^2}
   \sum_{\ir=1}^p w^{6,\ir}\mathrm{d}w^{5,\ir}+ 4 \, \mathrm{d}w^4\right) &\null & \null&\nonumber\\
  e^{5,\ir} &\,=\,  e^{-w^1} \left(\frac{1}{\sqrt{2}} \, e^{w^2} w^{6,\ir} \mathrm{d}w^3+ \,\mathrm{d}w^{5,\ir}\right) &\quad ; \quad & 
  &\nonumber\\
  e^{6,\ir} &\,=w^{6,\ir} \mathrm{d}w^2+\mathrm{d}w^{6,\ir}&\quad ; \quad & 
  &
\end{alignat}
\subsubsection{Conversion of L$(-1,p)$ solvable coordinates into CV solvable ones}
Relying on eq.s(\ref{econdor}) we easily obtain the conversion of the solvable coordinates of L$(-1,p)$ into the solvable ones of the corresponding CV manifold:
\begin{alignat}{4}
\label{luchang}
  \varphi &\,=\, -2 \sqrt{\frac{2}{3}} w^2&\null & \null&\nonumber\\
  \psi &\,=\,w^1 \, - \, \ft 13 \,w^2 &\null & \null &\nonumber\\
  x^1 &\,=\, - w^3 &\null & \null&\nonumber\\
   x^n &\,=\, 2 w^4 &\null & \null&\nonumber\\
 x^{1+\ir} &\,=\, w^{5,\ir} &\quad  \quad &
 &\nonumber\\
 \chi^\ir &\,= e^{w^2} \,w^{6,\ir} &\quad ; \quad & \ir=1,\dots,p \,.&
\end{alignat}
Relying on the above coordinate correspondence, in appendix \ref{trinop} we derive the explicit $\sym(6+3p,\mathbb{R})$ representation of the solvable generators $T_A =\{T_1,\ldots, T_4,T_{5,\ir},T_{6,\ir}\}$ dual to the $1$-forms $e^A$ satisfying the Maurer Cartan equations (\ref{maurocartus}) of which, in \cite{secondtemperature}, we provided the vector representation inside the fundamental irrep of $\mathrm{SO}(2,2+p)$ and its uploading to the defining irrep  of the symplectic group $\Sp(8+2p,\mathbb{R})$. Our reader can find the explicit $\sym(6+3p,\mathbb{R})$ representation of the $T_A$
in eq.s(\ref{aenadumgenetrix}-\ref{Bfin}); the latter are presented in the $p=3$ case but the generalization to all values of $p$ is obvious and straightforward.
\subsection{The K\"ahler $2$-form of L$(-1,p)$}
Next we consider the K\"ahler 2-form of the manifold L$(-1,p)$.
In complex coordinates, for any K\"ahler manifold one sets:
\begin{equation}\label{kalloforma}
  \mathbb{K} \, = \, \ft{\rmi}{2} \, \partial_\ial\partial_{\bar{\jbe}}\mathcal{K} \, \mathrm{d}z^\ial \wedge \mathrm{d}\bar{z}^{\bar{\jbe}}\,.
\end{equation}
For all those cases, as ours (see eq.(\ref{genKpotyonly})), where the K\"ahler potential $\mathcal{K}= \mathcal{K}(\mathbf{y})$ depends only from the imaginary parts $y^i$ ($i=1,\dots,n$) (see eq.(\ref{zetaspeciale})),the K\"ahler 2-form $\mathbb{K}$ reduces to:
\begin{equation}\label{kalloformaHes}
  \mathbb{K} \, = \, \sum_{\ial=1}^n \, \sum_{\jbe=1}^n \rmd x^\ial \wedge\rmd y^\jbe \frac{\partial^2\mathcal{K}(\mathbf{y})}{\partial y^\ial \partial y^\jbe}\,.
\end{equation}
Utilizing first eq.s (\ref{ytoLm1solv}) to express the $y^\ial$ in terms of L$(-1,p)$ solvable coordinates and then eq.(\ref{luchang}) to express the latter in terms of $\mathbf{w}$-coordinates we obtain:
\begin{equation}\label{krano}
  \mathbb{K} \, = \, K_{AB}(\mathbf{w}) \, \mathrm{d}w^A \wedge \mathrm{d}w^B\,, \qquad  A,B=1,\ldots ,2n\,,
\end{equation}
where the coefficients $K_{AB}(\mathbf{w}) $ are precisely determined functions of the coordinates $\mathbf{w}$ that we do not write here in order to save space and reader's time: indeed its explicit form is not particularly interesting. What is interesting is the form of $\mathbb{K}$ when in eq.(\ref{krano}) we replace the differentials of the solvable coordinates in terms of the left-invariant $1$-forms $e^A$. Indeed
we have the following inverse formula of eq.(\ref{econdor}) that expresses the differentials of the solvable coordinates $\mathbf{w}$ as linear combination of the left invariant $1$-forms $e^A$ with coordinate depending coefficients:
\begin{alignat}{4}
\label{rodnoce}
 \mathrm{d}w^1  &\,=\, e^1 &\null & \null&\nonumber\\
 \mathrm{d}w^2 &\,=\, e^2 &\null & \null &\nonumber\\
 \mathrm{d}w^3 &\,=\, e^{w^1-w^2}\, e^{3}&\null & \null&\nonumber\\
 \mathrm{d}w^4 &\,=\, \, e^{w^1+w^2} \left(-\frac{1}{4} e^3\, \sum_{\ir=1}^p (w^{6,\ir})^2\,+\frac{1}{\sqrt{2}}\,
 \sum_{\ir=1}^p e^{5,\ir}\, w^{6,\ir}+\, e^4\right) &\null & \null&\nonumber\\
  \mathrm{d}w^{5,\ir} &\,=\,e^{w_1} \left( -\frac{1}{\sqrt{2}} \,e^3 \, w^{6,\ir}+\,e^{5,\ir}\right) &\quad ; \quad & 
  &\nonumber\\
  \mathrm{d}w^{6,\ir} &\,=\,-\, w^{6,\ir}\, e^2\, + \, e^{6,\ir}.&\quad ; \quad & 
  &
\end{alignat}
Inserting eq.(\ref{rodnoce}) into eq.(\ref{krano}) we obtain:
\begin{equation}\label{Kn2}
   \mathbb{K} \, = \,2 \,\left(-\,e^1\wedge e^3 + e^1 \wedge e^4 + e^2 \wedge e^3 + e^2\wedge e^4\right) \, - \, \sqrt{2} \, \sum_{\ir=1}^{p} e^{5,\ir} \wedge e^{6,\ir} \,,
\end{equation}
which is exactly reproduced by the matrix $\mathcal{K}_{AB}$ anticipated in eq.(\ref{carbonero}).
\par
In order to clarify the correct interpretation of the above formula holding true for all values of $p$ we present below the
explicit case $p=3$.
\begin{equation}\label{rotKq3}
  \mathcal{K}_{p=3}\, = \, \left(
\begin{array}{cccc||ccc|ccc}
 0 & 0 & -1 & 1 & 0 & 0 & 0 & 0 & 0 & 0 \\
 0 & 0 & 1 & 1 & 0 & 0 & 0 & 0 & 0 & 0 \\
 1 & -1 & 0 & 0 & 0 & 0 & 0 & 0 & 0 & 0 \\
 -1 & -1 & 0 & 0 & 0 & 0 & 0 & 0 & 0 & 0 \\
 \hline\hline
 0 & 0 & 0 & 0 & 0 & 0 & 0 & -\frac{1}{\sqrt{2}} & 0 & 0 \\
 0 & 0 & 0 & 0 & 0 & 0 & 0 & 0 & -\frac{1}{\sqrt{2}} & 0 \\
 0 & 0 & 0 & 0 & 0 & 0 & 0 & 0 & 0 & -\frac{1}{\sqrt{2}} \\
 \hline
 0 & 0 & 0 & 0 & \frac{1}{\sqrt{2}} & 0 & 0 & 0 & 0 & 0 \\
 0 & 0 & 0 & 0 & 0 & \frac{1}{\sqrt{2}} & 0 & 0 & 0 & 0 \\
 0 & 0 & 0 & 0 & 0 & 0 & \frac{1}{\sqrt{2}} & 0 & 0 & 0 \\
\end{array}
\right)
\end{equation}
\section{The full isometry group of L$(-1,p)$}
\label{kallergenerator}
Making reference to appendix \ref{classmates} and to the above and there quoted literature on the classification of Special Homogeneous Geometries, it is known that the full isometry group of the Special K\"ahler Manifold L$(-1,p)$ is not exhausted by the solvable group $\mathcal{S}_{\mathrm{L}(-1,p)}$ semidirect product with the Paint Group $\mathrm{G_{Paint}} \, = \, \mathrm{SO}(p)$, that, by definition, is the outer automorphism group of $\mathcal{S}_{\mathrm{L}(-1,p)}=\mathcal{S}_{2,2+p}$. Indeed we know that there is an extra generator named $a_M$ in Fig.\ref{tbl:isohomvsk} of appendix \ref{classmates}, whose gradings, translated in the Cartan basis as defined by the Maurer Cartan equations (\ref{maurocartus}) namely with respect
to $e^1,e^2$ are $(-1,+1)$, while those of
the generator $T_3$ dual to $e^3$ are $(1,-1)$. Let us name $L_{-}$ this extra generator and rename
\begin{equation}\label{gutenabend}
  L_+ \equiv T_3 \quad ; \quad L_0\equiv \ft 12 \left(T_1-T_2\right)
\end{equation}
The idea, inspired by the results of the old classification papers \cite{productproof,SKGaggio3,SKGaggio2,SKGaggio1,specHomgeoA2,specHomgeoA1,vandersuppa,Ferrara:1989py,deWit:1995tf,Lauria:2020rhc} and in particular by fig.\ref{tbl:isohomvsk} of the appendix, taken from \cite{vandersuppa,deWit:1995tf}, is the following:
\begin{enumerate}
  \item The generators $L_{\pm},L_0$ close a standard $\slal(2,\mathbb{R})$ Lie algebra, namely:
      \begin{equation}\label{robinia1}
        \left[L_0 , L_{\pm}\right] \, = \, \pm L_\pm \quad ; \quad \left[L_+ , L_{-}\right] \, = \,2 L_0
      \end{equation}
  \item The painted generators $T_{5,\ir},T_{6,\ir}$ form $p$ doublet  fundamental representations of such a $\slal(2,\mathbb{R})$ Lie algebra.
      In addition to the already existing commutation relations of $T_{5,\ir},T_{6,\ir}$ with the $L_+,L_0$ generators (that are simply renamings of $T_3$ and $\ft 12 \left(T_1-T_2\right)$) such a hypothesis requires the commutation relations:
     \begin{equation}\label{robinia2}
        \left[L_- , T_{5,\ir}\right] \, = \, \sqrt{2} \, T_{6,\ir}  \quad ; \quad \left[L_- , T_{6,\ir}\right] \, = \, 0
      \end{equation}
\end{enumerate}
That the above two hypotheses are viable and that the addition
of an $L_-$ generator with the above commutation relations closes
an abstract Lie algebra (necessarily non-semisimple) can be checked directly via a deformation of the Maurer Cartan equations (\ref{maurocartus}) through the addition of a new
$1$-form $\Psi_-$ as it follows:
\begin{equation}\label{maurocartus2}
  \begin{array}{lcl}
 \mathrm{d}e^1-\frac{1}{4} e^3\wedge \Psi_{-}&=& 0 \\
 \mathrm{d}e^2+\frac{1}{4} e^3\wedge \Psi_{-}&=& 0 \\
 \mathrm{d}e^3+e^1\wedge e^3-e^2\wedge e^3&=& 0 \\
 \mathrm{d}e^4+e^1\wedge e^4+e^2\wedge e^4-\frac{1}{\sqrt{2}}\,\sum_{\ir=1}^p \,e^{5,\ir}\wedge e^{6,\ir}
 &=& 0 \\
  \mathrm{d}e^{5,\ir}+e^1\wedge e^{5,\ir}\,+\frac{1}{\sqrt{2}}\, e^3\wedge e^{6,\ir}&=& 0\\
 \mathrm{d}e^{6,\ir}+e^2\wedge e^{6,\ir}\,-\,\frac{1}{2\sqrt{2}}\, \Psi_{-}\wedge e^{5,\ir}&=& 0 \\
 \mathrm{d}\Psi_{-}-e^1\wedge \Psi_{-}+e^2\wedge \Psi_{-}&=& 0 \,.\\
\end{array}
\end{equation}
By taking the exterior derivative of  equations (\ref{maurocartus2}) and then replacing for $\mathrm{d}e^A,\rmd\Psi_{-}$ the expressions provided by the same equations, we verify that the differential system (\ref{maurocartus2}) is self-consistent, namely eq.s (\ref{maurocartus2}) define an  existing abstract Lie algebra. On the other hand if we set $\Psi_- \to 0$ the Maurer Cartan equations (\ref{maurocartus2}) contract to the previously established ones (\ref{maurocartus}), that, as we know, are consistent. That means that we should be able to find in the carrier space of each representation of the solvable Lie algebra (\ref{maurocartus}) an appropriate matrix $L_-$ that closes with
the solvable generators $T_A$ the Lie algebra defined by the Maurer Cartan equations (\ref{maurocartus2}).
\subsection{Determination of $L_-$ inside the symplectic Lie algebra $\sym(6+2p,\mathbb{R})$}
\label{sp6extragen}
We were able to determine the matrix $L_-$ \textbf{uniquely} as a symplectic matrix belonging to the Lie Algebra $\sym(6+2p,\mathbb{R})$. Starting from the most general element
of $\sym(6+2p,\mathbb{R})$ as given in eq.s(\ref{gensymmatr}-\ref{croupier}) of appendix \ref{simplettando}:
\begin{equation}\label{pisano}
  L_- \, = \, \Lambda
\end{equation}
and from the generators of the solvable Lie algebra  as given
in eq.s (\ref{aenadumgenetrix}-\ref{Bfin}) we imposed the commutation relations:
\begin{eqnarray}\label{boacostrictor}
  \left[L_-\,,\, T_{5,\ir}\right] &=& \sqrt{2} \, T_{6,\ir} 
  \nonumber \\
  \left[L_-\,,\, T_{6,\ir}\right] &=& 0 
  \nonumber \\
  \left[L_0\,,\, L_-\right] &=& - \, L_- \nonumber\\
  \left[L_+\,,\, L_-\right] &=& 2 \, L_0\,,
\end{eqnarray}
where $L_+,L_0$ are identified above in eq.(\ref{gutenabend}) in terms of the solvable generators. The plethora of parameters, whose number is given in eq.(\ref{cratere}), are all completely fixed by the constraints (\ref{boacostrictor}) and the solution is given by the following matrix in the case $p=3$: (again in the basis  (\ref{ololo}) with $\Lambda ,\Sigma =0,1,\ir,n$)
\begin{equation}\label{Lminusmatrix}
  L_- \, = \, \left(
\begin{array}{cc|ccc|c||cc|ccc|c}
 0 & 0 & 0 & 0 & 0 & 0 & 0 & 0 & 0 & 0 & 0 & 0 \\
 -1 & 0 & 0 & 0 & 0 & 0 & 0 & 0 & 0 & 0 & 0 & 0 \\
 \hline
 0 & 0 & 0 & 0 & 0 & 0 & 0 & 0 & 0 & 0 & 0 & 0 \\
 0 & 0 & 0 & 0 & 0 & 0 & 0 & 0 & 0 & 0 & 0 & 0 \\
 0 & 0 & 0 & 0 & 0 & 0 & 0 & 0 & 0 & 0 & 0 & 0 \\
 \hline
 0 & 0 & 0 & 0 & 0 & 0 & 0 & 0 & 0 & 0 & 0 & 0 \\
 \hline\hline
 0 & 0 & 0 & 0 & 0 & 0 & 0 & 1 & 0 & 0 & 0 & 0 \\
 0 & 0 & 0 & 0 & 0 & -1 & 0 & 0 & 0 & 0 & 0 & 0 \\
 \hline
 0 & 0 & \frac{1}{2} & 0 & 0 & 0 & 0 & 0 & 0 & 0 & 0 & 0 \\
 0 & 0 & 0 & \frac{1}{2} & 0 & 0 & 0 & 0 & 0 & 0 & 0 & 0 \\
 0 & 0 & 0 & 0 & \frac{1}{2} & 0 & 0 & 0 & 0 & 0 & 0 & 0 \\
 \hline
 0 & -1 & 0 & 0 & 0 & 0 & 0 & 0 & 0 & 0 & 0 & 0 \\
\end{array}
\right)
\end{equation}
It is straightforward to generalise the result for $p=3$ displayed above to all values of $p$.
\par
Instead of the non-compact generator $L_-$, as addition to the solvable Lie algebra we can consider the compact generator:
\begin{equation}\label{camisablanca}
 \mathrm{ H_L} \, \equiv \, L_+ - L_-
\end{equation}
whose explicit expression is the following:
\begin{equation}\label{camisastretta}
  \mathrm{ H_L}\, = \,\left(
\begin{array}{cc|ccc|c||cc|ccc|c}
 0 & -2 & 0 & 0 & 0 & 0 & 0 & 0 & 0 & 0 & 0 & 0 \\
 1 & 0 & 0 & 0 & 0 & 0 & 0 & 0 & 0 & 0 & 0 & -2 \\
 \hline
 0 & 0 & 0 & 0 & 0 & 0 & 0 & 0 & 2 & 0 & 0 & 0 \\
 0 & 0 & 0 & 0 & 0 & 0 & 0 & 0 & 0 & 2 & 0 & 0 \\
 0 & 0 & 0 & 0 & 0 & 0 & 0 & 0 & 0 & 0 & 2 & 0 \\
 \hline
 0 & 0 & 0 & 0 & 0 & 0 & 0 & -2 & 0 & 0 & 0 & 0 \\
 \hline\hline
 0 & 0 & 0 & 0 & 0 & 0 & 0 & -1 & 0 & 0 & 0 & 0 \\
 0 & 0 & 0 & 0 & 0 & 1 & 2 & 0 & 0 & 0 & 0 & 0 \\
 \hline
 0 & 0 & -\frac{1}{2} & 0 & 0 & 0 & 0 & 0 & 0 & 0 & 0 & 0 \\
 0 & 0 & 0 & -\frac{1}{2} & 0 & 0 & 0 & 0 & 0 & 0 & 0 & 0 \\
 0 & 0 & 0 & 0 & -\frac{1}{2} & 0 & 0 & 0 & 0 & 0 & 0 & 0 \\
 \hline
 0 & 1 & 0 & 0 & 0 & 0 & 0 & 0 & 0 & 0 & 0 & 0 \\
\end{array}
\right)
\end{equation}
From equation (\ref{camisastretta}) we see that the generator $\mathrm{H_L}$ is indeed a compact operator since it admits only purely imaginary eigenvalues but it is not an element of the maximal compact subalgebra $\uu(3)$ as defined in appendix \ref{simplettando}. In particular we can compare eq.(\ref{camisastretta}) with eq. (\ref{normaloide}) and we see that in no way the generator $\mathrm{H_L}$ can be inserted in the normalizer of the Paint Group in $\uu(3)$. Neither can help the idea of choosing a different combination of $L_+$ and $L_-$:
$$ \mathrm{H}^\prime \, = \, L_+ - \alpha L_- \quad ; \quad \alpha >0$$
For no value of $\alpha$, the generator $\mathrm{H}^\prime$ can be fitted into the normalizer. We stressed the uniqueness of the solution precisely for this reason. Such uniqueness excludes that the unique $\uu(1)$  isometry generator of L$(-1,p)$ can belong to the maximal compact subalgebra of $\sym(6+2p,\mathbb{R})$, just as it does not belong to the compact isometry algebra of the Calabi-Vesentini manifolds.
\par
Finally we can also exclude that the constant components of K\"ahler $2$-form as displayed in eq.(\ref{rotKq3}) or (\ref{carbonero}) be related to the adjoint action of $\mathrm{H_L}$ as it is instead the case in all K\"ahler symmetric spaces. One can do detailed analysis but the most evident obstacle is that $\mathrm{H_L}$ commutes with $T_4$ so that the adjoint action yields a matrix of rank inferior to the maximal one and hence can in no way provide the components of the K\"ahler $2$-form that is of maximal rank.
\subsection{Determination of $L_-$ inside the pseudo-orthogonal Lie algebra $\so(2,2+p)$}
\label{pseudortho}
As we anticipated in the introduction (subsection \ref{ciufolino}), following a logic completely identical to that employed in the previous section \ref{sp6extragen}, given the explicit realization of the $Solv_{[2,2+p]}$ generators as matrices belonging to the $\so(2,2+p)$ Lie algebra (see section
\ref{sciogliCV} and references there quoted) one can try to determine an element $L_-\in\so(2,2+p)$ of the same algebra and in the same basis that satisfies the commutation relations (\ref{robinia1}-\ref{robinia2}). A posteriori not too surprisingly, we find just one unique solution of such a problem, which is the following (as usual we utilize the $p=3$ case as an illustration that immediately reveals the general pattern): (basis as in (\ref{etatdefi}))
\begin{equation}\label{Lminusortho}
  L_-^{\so} \, =\, \left(
\begin{array}{cc|ccc|cc}
 0 & 0 & 0 & 0 & 0 & 0 & 0 \\
 \sqrt{2} & 0 & 0 & 0 & 0 & 0 & 0 \\
 \hline
 0 & 0 & 0 & 0 & 0 & 0 & 0 \\
 0 & 0 & 0 & 0 & 0 & 0 & 0 \\
 0 & 0 & 0 & 0 & 0 & 0 & 0 \\
 \hline
 0 & 0 & 0 & 0 & 0 & 0 & 0 \\
 0 & 0 & 0 & 0 & 0 & -\sqrt{2} & 0 \\
\end{array}
\right)
\end{equation}
The lines dividing the matrix (\ref{Lminusortho}) into sub-blocks follow the pattern established in eq.(\ref{etatdefi}) by the invariant metric $\eta_t$ of the triangular basis (see \cite{pgtstheory}). The set of solvable generators plus $L_-$ closes a subalgebra of $\so(2,2+p)$ obviously isomorphic to the algebra defined by the Maurer Cartan equations (\ref{maurocartus2}). Inside such an algebra we have the generator $\mathrm{H_L}$ defined in eq.(\ref{camisablanca}) whose representation inside $\so(2,2+p)$ is given by the following matrix:
\begin{equation}\label{soHL}
  \mathrm{H_L}^{\so} \, = \, \left(
\begin{array}{cc|ccc|cc}
 0 & \frac{1}{\sqrt{2}} & 0 & 0 & 0 & 0 & 0 \\
 -\sqrt{2} & 0 & 0 & 0 & 0 & 0 & 0 \\
 \hline
 0 & 0 & 0 & 0 & 0 & 0 & 0 \\
 0 & 0 & 0 & 0 & 0 & 0 & 0 \\
 0 & 0 & 0 & 0 & 0 & 0 & 0 \\
 \hline
 0 & 0 & 0 & 0 & 0 & 0 & -\frac{1}{\sqrt{2}} \\
 0 & 0 & 0 & 0 & 0 & \sqrt{2} & 0 \\
\end{array}
\right)
\end{equation}
The matrix $\mathrm{H_L}^{\so}$ generates a $\mathrm{U(1)}$ subgroup of $\mathrm{SO}(2,2+p)$ since it has two pairs of complex conjugate eigenvalues $\pm i$ and $p$ eigenvalues $0$. Yet $\mathrm{H_L}^{\so}$ does not belong to the canonical compact subalgebra $\mathbb{H}_c^{CV} \subset \so(2,2+p)$. Obviously with some appropriate $\mathrm{SO}(2,2+p)$ transformation we might rotate $\mathrm{H_L}^{\so}$ into the canonical compact subalgebra, yet the same transformation has to be applied to the $\mathbb{K}^{CV}$ coset  generator subspace whose image will be orthogonal to another $\mathbb{H}_c^\prime$ subalgebra, leaving the conclusion unchanged: \textit{$\mathrm{H_L}^{\so}$ does not belong to the isotropy subalgebra of the Calabi-Vesentini symmetric space.}
The above conclusion refers to the metric properties that are obviously different for L$(-1,p)$ and $\mathcal{M}^{[2,p]}_{CV}$
and perfectly consistent with the matter of fact that the two different metrics are imposed on the same solvable Lie Group $\mathcal{S}_{[2,2+p]}$.
\par
A further clarifying remark is provided by the decomposition of the operator $\mathrm{H_L}^{\so}$ in the orthogonally decomposed
basis $\mathbb{K}\oplus \mathbb{H}$ as explicitly described in \cite{secondtemperature}. We have:
\begin{equation}\label{HKdeco}
  \mathrm{H_L}^{\so} \, = \, - \frac{1}{\sqrt{2}} \, \mathrm{K_3}\, +\, \frac{3}{2} \left(\mathrm{H_1} + \mathrm{H_2}\right)
\end{equation}
Calculating the adjoint representation of $\mathrm{H_L}^{\so}$ inside $\so(2,2+p)$ we see that it mixes the $\mathbb{K}$-subspace with the $\mathbb{H}$-subalgebra.
\section{The $\mathrm{H_L}$ transformation on solvable coordinates}
\label{baldacchino}
Apart from the Paint Group transformations and the solvable Lie group translations the only other isometry of the L$(-1,p)$ manifold is that generated by either $L_-$ or $\mathrm{H_L}^{\so}$. Since the latter is compact it is the most interesting one for use in Neural Network architectures. The question is how to calculate its action on the solvable coordinates. Such a question can be answered in a simple way if we consider the solvable group as a coset manifold where the numerator group is the isometry group $\mathrm{Iso}_{\mathrm{L}(-1,p)}$
and the denominator subgroup is the translation group generated by $L_-$ that we can name $T_-\,\sim\,\exp[L_-]$. In other words we consider the map sequence:
\begin{equation}\label{sturmundrang}
  Solv_{[2,2+p]} \, \stackrel{\iota}{\hookrightarrow} \,\mathrm{Iso}_{\mathrm{L}(-1,p)} \, \stackrel{\pi}{\longrightarrow} \, \frac{\mathrm{Iso}_{\mathrm{L}(-1,p)}}{T_-} \, \simeq \, Solv_{[2,2+p]}
\end{equation}
where $\iota$ is the injection map and $\pi$ the quotient map with respect to the subgroup.
\par
Relying on this vision, the solvable group element (we always use the case $p=3$ as an instrument to obtain the general result):
\begin{eqnarray}\label{francioso}
 &&\mathbb{L}(w) \, = \,\nonumber\\
 &&
{\fontsize{6.0}{6.3}
\left(
\begin{array}{ll|lll|ll}
 e^{w^1} & \frac{e^{w^2} w^3}{\sqrt{2}} & \frac{1}{2} \left(\sqrt{2} w^{5,1}\right. & \frac{1}{2} \left(\sqrt{2} w^{5,2}\right. &
   \frac{1}{2} \left(\sqrt{2} w^{5,3}\right. & \frac{1}{8} e^{-w^2}
   \left(-4 w^{5,1} w^{6,1}-4 w^{5,2} w^{6,2}-4 w^{5,3}
   w^{6,3}\right. & -\frac{1}{4} e^{-w^1}
   \left((w^{5,1})^2+(w^{5,2})^2\right. \\
   \null&\null&\left.+w^3
   w^{6,1}\right)&\left.+w^3 w^{6,2}\right)&\left.+w^3 w^{6,3}\right)&\left.-\sqrt{2} w^3 \left((w^{6,1})^2+(w^{6,2})^2+(w^{6,3})^2\right)+4
   \sqrt{2} w^4\right)&\left.+(w^{5,3})^2+2 w^3 w^4\right) \\
 0 & e^{w^2} & \frac{w^{6,1}}{\sqrt{2}} & \frac{w^{6,2}}{\sqrt{2}} &
   \frac{w^{6,3}}{\sqrt{2}} & -\frac{1}{4} e^{-w^2}
   \left((w^{6,1})^2+(w^{6,2})^2+(w^{6,3})^2\right) & -\frac{e^{-w^1} w^4}{\sqrt{2}} \\
   \hline
 0 & 0 & 1 & 0 & 0 & -\frac{e^{-w^2} w^{6,1}}{\sqrt{2}} & -\frac{e^{-w^1}
   w^{5,1}}{\sqrt{2}} \\
 0 & 0 & 0 & 1 & 0 & -\frac{e^{-w^2} w^{6,2}}{\sqrt{2}} & -\frac{e^{-w^1}
   w^{5,2}}{\sqrt{2}} \\
 0 & 0 & 0 & 0 & 1 & -\frac{e^{-w^2} w^{6,3}}{\sqrt{2}} & -\frac{e^{-w^1}
   w^{5,3}}{\sqrt{2}} \\
   \hline
 0 & 0 & 0 & 0 & 0 & e^{-w^2} & -\frac{e^{-w^1} w^3}{\sqrt{2}} \\
 0 & 0 & 0 & 0 & 0 & 0 & e^{-w^1} \\
\end{array}
\right)} \nonumber\\
\end{eqnarray}
can be regarded as a coset representative of the coset $\frac{\mathrm{Iso}_{\mathrm{L}(-1,p)}}{T_-}$. As it happens for all coset manifolds of type $\mathrm{G/H}$, given a coset representative $\mathbb{L}(\mathbf{w})$ that is an element of the group $\mathrm{G}$ depending on as many parameters $\mathbf{w}$ as is the dimension of $\mathrm{G/H}$ in such a way that $\mathbb{L}(\mathbf{w})$ and $\mathbb{L}(\mathbf{w}^\prime )$ are $\mathrm{H}$-equivalent \textbf{if and only if} $\mathbf{w}\, = \,\mathbf{w}^\prime $ and given, furthermore any element $g\in \mathrm{G}$, the action of $g$ on the parameters
$\mathbf{w}$ is obtained by setting:
\begin{equation}\label{cosettando}
  g \, \mathbb{L}(\mathbf{w}) \, h(\mathbf{w},g) \, = \, \mathbb{L}(g[\mathbf{w}])
\end{equation}
where $h(\mathbf{w},g)\in \mathrm{H}$, named \textbf{the compensator} is an element depending both on the parameters $\mathbf{w}$ and the group element $g$ such that the object $g \, \mathbb{L}(\mathbf{w}) \, h(\mathbf{w},g)$ falls in the parameterization $\mathbb{L}(\mathbf{u})$ for new parameters
$\mathbf{u}$ that depend on $\mathbf{w}$ and $g$.
\par
In our case, having a one-dimensional isometry group and a one dimensional subgroup $T_-$ the matter is rather simple.
For the compact group element we have:
\begin{equation}\label{gelement}
  g\, \equiv \, \exp[\theta \mathrm{H_L}] \, =\, \left(
\begin{array}{cc|ccc|cc}
 \cos (\theta ) & \frac{\sin (\theta )}{\sqrt{2}} & 0 & 0 & 0 & 0 & 0 \\
 -\sqrt{2} \sin (\theta ) & \cos (\theta ) & 0 & 0 & 0 & 0 & 0 \\
 \hline
 0 & 0 & 1 & 0 & 0 & 0 & 0 \\
 0 & 0 & 0 & 1 & 0 & 0 & 0 \\
 0 & 0 & 0 & 0 & 1 & 0 & 0 \\
 \hline
 0 & 0 & 0 & 0 & 0 & \cos (\theta ) & -\frac{\sin (\theta )}{\sqrt{2}} \\
 0 & 0 & 0 & 0 & 0 & \sqrt{2} \sin (\theta ) & \cos (\theta ) \\
\end{array}
\right)
\end{equation}
while the compensator
\begin{equation}\label{superkalifragilisti}
  h(\mathit{f}) \, =\,\exp[\mathit{f} \, \mathrm{H_L}] \, = \, \left(
\begin{array}{cc|ccc|cc}
 1 & 0 & 0 & 0 & 0 & 0 & 0 \\
 \sqrt{2}\, \mathit{f} & 1 & 0 & 0 & 0 & 0 & 0 \\
 \hline
 0 & 0 & 1 & 0 & 0 & 0 & 0 \\
 0 & 0 & 0 & 1 & 0 & 0 & 0 \\
 0 & 0 & 0 & 0 & 1 & 0 & 0 \\
 \hline
 0 & 0 & 0 & 0 & 0 & 1 & 0 \\
 0 & 0 & 0 & 0 & 0 & -\sqrt{2}\, \mathit{f}  & 1 \\
\end{array}
\right)
\end{equation}
is a lower triangular matrix.
\par
The compensator parameter is immediately fixed by the request that  the object $g \, \mathbb{L}(\mathbf{w}) \, h(\mathit{f})$ should be upper triangular. Indeed we find:
\begin{equation}\label{ilmiocompenso}
  \mathit{f} \, = \, \frac{e^{w^1-w^2}}{\cot (\theta )-w^3}
\end{equation}
After the substitution (\ref{ilmiocompenso}) one finds that
\begin{equation}\label{gromello}
  g \, \mathbb{L}(\mathbf{w}) \, h\left(\frac{e^{w^1-w^2}}{\cot (\theta )-w^3}\right) \, = \, \mathbb{L}(\mathbf{u})
\end{equation}
where the new solvable coordinates $\mathbf{u}$ are related to the old ones $\mathbf{w}$ in the following way:
\begin{alignat}{4}\label{berlinoest}
  u^1 &\, = \,& w^1 -\log \left[\cos (\theta )-w^3 \sin (\theta )\right] &\quad;\quad &\null\nonumber \\
  u^2 &\, = \,& w^2 + \log \left[\cos (\theta )-w^3 \sin (\theta )\right] &\quad;\quad &\null\nonumber \\
  u^3 &\, = \,& \frac{\sin (\theta )+w^3 \cos (\theta )}{\cos (\theta )-w^3 \sin (\theta )}&\quad;\quad &\null\nonumber \\
  u^4 &\, = \,& w^4+\frac{\left(\sum_{\ir=1}^p (w^{5,\ir})^2\right) \sin (\theta )}{2 \left(w^3 \sin
   (\theta )-\cos (\theta )\right)} &\quad;\quad &\null\nonumber\\
   u^{5,\ir}&\, = \,& \frac{w^{5,\ir}}{\cos (\theta )-w^3 \sin (\theta )} &\quad;\quad & 
   \nonumber\\
   u^{6,\ir}&\, = \,&w^{6,\ir} \cos (\theta )-\left(\sqrt{2}\,w^{5,\ir}+w^3 w^{6,\ir}\right) \sin (\theta )&\quad.\quad & 
   \end{alignat}
   The transformation (\ref{berlinoest}) is the main new tool in designing the architecture of $r=2$ Cartan Neural Networks contributed by the intriguing and unexpected identification of the solvable Lie group of CV manifolds with that of the L$(-1,p)$ Special K\"ahler manifolds.
\section{Conclusions}
As anticipated in the introduction, in this paper we have proved that the Special K\"ahler Manifold L$(-1,p)$ is an Alekseevskyan space on the same solvable group manifold $\mathcal{S}_{[2,2+p]}$
that is the underlying manifold of the Calabi-Vesentini symmetric space $\mathrm{SO}(2,2+p)/\mathrm{SO}(2)\times \mathrm{SO}(2+p)$.
The L$(-1,p)$ structure just arises by defining a quadratic form on the solvable Lie Algebra $Solv_{[2,2+p]}$ that is different from the one corresponding to the symmetric space and leads to a smaller isometry group.
We have also analysed in depth the conversion from the special coordinates utilized in the cubic polynomial setup and the solvable coordinates naturally arising in the treatment of CV manifolds. Particular attention was devoted to the explicit construction of the matrix realization of the isometry generators in the symplectic group associated with Special K\"ahler Geometry. Finally we were able to show that the L$(-1,p)$ manifold identified with its solvable Lie group and the completion of the latter to the full isometry group $\mathrm{Iso}_{\mathrm{L}(-1,p)}$ can be directly realized inside the isometry group $\mathrm{SO}(2,2+p)$ of CV manifolds. The catch of the matter is that the only compact generator of $\mathrm{Iso}_{\mathrm{L}(-1,p)}$ commuting with the Paint Group does not belong to the compact subalgebra of the CV symmetric space. This makes the discovered identification relevant for Neural Networks, because the compact isometry associated with the L$(-1,p)$ interpretation is not available in the symmetric space interpretation and therefore provides an additional overall non linear transformation of the solvable coordinates that can contribute to increase the expressivity of the net.
\par
Apart from its use in CaNNs architecture our new result has an intrinsic interest for the theory of Supergravity special geometries since it reveals the close connection between the familiar CV manifolds and the homogeneous non symmetric space L$(-1,p)$ that can be constructed directly inside $\mathrm{SO}(2,2+p)$ without the need to use the more complicated $\sym(6+2p,\mathbb{R})$ setup, just introducing a different quadratic form on the solvable Lie algebra $Solv_{[2,2+p]}$.
\par
There is an ample set of questions that are generated by this unexpected identification, the answers to which might be relevant both for the fundamental mathematical theory and for ML applications. Let us briefly enumerate them:
\begin{enumerate}
  \item Since on the very same manifold, described by the very same coordinates of its unique open chart (Cartan-Hadamard theorem) there are two K\"ahler metrics and hence two different families of geodesics, what is the relation between the two geodesics joining any given
      pair of points, the CV one and the L$(-1,p)$ one?
  \item Similarly on the same manifold and within the same coordinate patch we have two K\"ahler 2-forms and hence two different Poissonian structures. What is the relation between them?
  \item The separator theory, vital for the construction of CaNN classification Networks is rooted in  CV geometry. Yet since a separator is a submanifold of the CV manifold it is necessarily also a submanifold of the corresponding L$(-1,p)$ space, which, apart from the metric structure, is the same variety. Is the CV separator also a separator in the L$(-1,p)$ interpretation and what is its algebraic status from the point of view of the much smaller isometry group $\mathrm{Iso}_{\mathrm{L}(-1,p)}$?
  \item Another aspect worth mentioning is that the Gibbs distributions on CV manifolds constructed and studied in \cite{secondtemperature,terzatemperatura} do not seem to have any obvious counterpart on L$(-1,p)$ spaces precisely because the  K\"ahler structure of the latter is not related to isometries and some of the compact generators associated with the definition of Souriau temperatures are missing. Yet the microscopic manifold is the same, although equipped with a different metric, so that a relation must exist and this poses another intriguing question liable to lead to some deeper understanding of the thermodynamical properties.
\end{enumerate}
\par
Finally a preliminary inspection of the list of special homogeneous geometries suggests that what we have discovered and analyzed in this paper might just be the tip of an iceberg since other identifications between different Special Manifolds at the level of the underlying solvable group manifold might exist.
The analysis of this possibility is postponed to a future publication.
\newpage
\appendix
\section{General form of the  $\sym(6+2p,\mathbb{R})$-Lie algebra elements.}
\label{simplettando}
As we recalled in section \ref{sommarioholosect}, every Special K\"ahler manifold is completely determined by a suitable holomorphic symplectic section $\Omega(z)$ that belongs to the fundamental representation of the Lie group $\Sp(2+2n,\mathbb{R})$, where $n$ is the complex dimension of the manifold. For homogeneous manifolds (whether symmetric or not) it follows that the generators of the Lie algebra of the motion group (the solvable Lie group in our case) admit a basic representation in the fundamental of $\sym(2+2n,\mathbb{R})$ namely in the fundamental of $\sym(6+2p,\mathbb{R})$ in our case. This is the rationale of the construction in section \ref{solvableinsp6}. In order to appreciate better that construction and to lay down the instruments used in particular in section \ref{kallergenerator}, in this appendix we summarize the structure of a generic element of the Lie algebra  $\sym(6+2p,\mathbb{R})$. As we do elsewhere in the paper we illustrate the general framework with the example $p=3$ that leads to $\sym(12,\mathbb{R})$.
\par
As in section \ref{sommarioholosect} let the invariant symplectic matrix be:
\begin{equation}\label{Cmatra}
  \mathbb{C}_s \, = \, \left(\begin{array}{c|c}
                              \mathbf{0}_{(3+p)\times(3+p)} & \mathbf{1}_{(3+p)\times(3+p)} \\
                               \hline
                               -\mathbf{1}_{(3+p)\times(3+p)} & \mathbf{0}_{(3+p)\times(3+p)}\,.
                             \end{array}
   \right)
\end{equation}
A generic element of $\sym(6+2p,\mathbb{R})$ is a matrix
\begin{equation}\label{gensymmatr}
  \Lambda \, = \, \left(\begin{array}{c|c}
                              A & B \\
                               \hline
                              C & D
                             \end{array}
   \right)
\end{equation}
that satisfies the constraint:
\begin{equation}\label{sympatconst}
  \Lambda^T \,  \mathbb{C}_s +  \mathbb{C}_s \, \Lambda \, = \, 0
\end{equation}
which implies the following relations among the $(3+p)\times(3+p)$ sub-blocks:
\begin{equation}\label{croupier}
  D \, = \, - A^T \quad ; \quad B\, = \, B^T \quad ; \quad C\, = \, C^T
\end{equation}
The dimension of the symplectic algebra is given:
\begin{equation}\label{cratere}
  \mathrm{dim}\,[\sym(6+2p,\mathbb{R})]\, = \, \ft 12 \,(6+2p)(7+2p)
\end{equation}
and we can easily check that this counting agrees with:
\begin{eqnarray}\label{vortice}
  \mathrm{dim}\left[\sym(6+2p,\mathbb{R})\right]& = & \mathrm{dim}[A] + \mathrm{dim}[B] + \mathrm{dim}[C] \nonumber\\
  \mathrm{dim}[A]& = & (3+p)^2 \nonumber\\
  \mathrm{dim}[B] & = & \ft 12 \, (3+p)(4+p) \nonumber\\
  \mathrm{dim}[C] & = & \ft 12 \, (3+p)(4+p)\,.
\end{eqnarray}
In the $p=3$ case we have:
\begin{equation}\label{12male}
  \mathrm{dim}\,\sym(12,\mathbb{R})\, = \, 78\,.
\end{equation}
The maximal compact subalgebra $\mathbb{H}_c$ of $\sym(6+2p,\mathbb{R})$ is
\begin{equation}\label{granarolo}
  \sym(6+2p,\mathbb{R})\, \supset \, \mathbb{H}_c \, \equiv \, \uu(3+p) \,  \cong \,\uu(1) \oplus \su(3+p)\,,
\end{equation}
which is realized as follows:
\begin{eqnarray}\label{gringo}
\su(3+p) \, \ni \, \Lambda & = & \left(\begin{array}{c|c}
\mathfrak{A} & \mathfrak{B} \\
\hline
-\, \mathfrak{B} & \mathfrak{A}
\end{array}\right )\nonumber\\
\mathfrak{A} & = & - \mathfrak{A}^T \quad\quad\quad\quad\quad\quad\,\, \text{antisymmetric}\nonumber\\
\mathfrak{B} & = & \mathfrak{B}^T, \quad \text{Tr}[\mathfrak{B}]\,=\, 0\quad \text{symmetric traceless,}
\end{eqnarray}
while the $\uu(1)$ subalgebra is simply generated by the symplectic matrix $\mathbb{C}_s$:
\begin{eqnarray}\label{stringo}
\uu(1) \, \ni \, \Lambda & = & \left(\begin{array}{c|c}
\mathbf{0} & b \mathbf{1} \\
\hline
-\, b \,\mathbf{1} & \mathbf{0}
\end{array}\right ) \, = \, b \, \mathbb{C}_s\,.
\end{eqnarray}
The quotient:
\begin{equation}\label{Siegelhalfplane}
  \mathbb{HS}_{3+p} \, \equiv \, \frac{\sym(6+2p,\mathbb{R})}{\uu(1) \oplus \su(3+p) }
\end{equation}
is a maximally split non-compact symmetric space, the \textit{Siegel half-plane}, whose dimension is:
\begin{equation}\label{dimmasiegela}
   \mathrm{dim}\left[\mathbb{HS}_{3+p}\right] \, = \, (3+p)(4+p)
\end{equation}
A standard conventional parameterization of the symmetric space is obtained by using the orthogonal decomposition:
\begin{equation}\label{ortodec}
  \sym(6+2p,\mathbb{R}) \, = \, \mathbb{K}_{Sieg} \oplus \uu(1) \oplus \su(3+p)
\end{equation}
where:
\begin{eqnarray}\label{fluberto}
  \mathbb{K}_{Sieg}\ni \Lambda & = & \left(\begin{array}{c|c}
\mathrm{A} & \mathrm{B} \\
\hline
\, \mathrm{B} & -\mathrm{A}^T
\end{array}\right )\nonumber\\
\mathrm{A} & = & \mathrm{A}^T \quad\quad\quad\,\, \text{symmetric}\nonumber\\
\mathrm{B} & = & \mathrm{B}^T\quad\quad\quad\,\, \text{symmetric.}
\end{eqnarray}
As usual the orthogonal subspace of Siegel half plane coset representatives $\mathbb{K}_{Sieg}$ does not close a subalgebra and leads to a non polynomial structure under exponentiation from the Lie algebra to the group. Yet, as it is true for all non-compact symmetric spaces, we can combine each $\mathbb{K}_{Sieg}$ matrix  with a matrix from the compact subalgebra and obtain the solvable Lie algebra that provides the solvable parameterization of the Siegel half plane.
Indeed the Siegel solvable Lie algebra is defined by the following conditions:
\begin{eqnarray}\label{solvsiegel}
  Solv^{Sieg}_{3+p} & = &\left\{\Lambda = \left(\begin{array}{c|c}
\mathrm{A}_{>} & \mathrm{B} \\
\hline
\, \mathbf{0} & -\mathrm{A}_{>}^T
\end{array}\right ) \, | \,A_{>}\,=\, \text{upper triangular}\, , \, \mathrm{B}=\mathrm{B}^T \, \, \text{symmetric}\right\}
\end{eqnarray}
It is easy to check that the defining conditions on the matrix $\Lambda$ are preserved by the commutator so that indeed $ Solv^{Sieg}_{3+p}$ is a Lie algebra. Furthermore one easily checks that the number of parameters is just equal to the dimension of the Siegel half-plane.
\par
It appears now evident that the choice of the solvable generators in section \ref{solvableinsp6} corresponds to an immersion of solvable Lie algebras:
\begin{equation}\label{immersion}
  Solv_{\mathrm{L}(-1,p)} \,\stackrel{\iota}{\longrightarrow} \, Solv^{Sieg}_{3+p}
\end{equation}
By exponentiation one comes to the conclusion that the Special Kahler manifold L$(-1,p)$ can be immersed into the Siegel half plane as a solvable subgroup of its metric equivalent solvable group.
\subsection{The Paint Algebra and its normalizer in $\uu(3)$}
An important item in our analysis of L$(-1,p)$ geometrical and algebraic structure is the Paint Lie Algebra and its position inside $\sym(6+2p,\mathbb{R})$. As announced we utilize the case $p=3$ and we display the matrix of a generic element of the Paint Lie algebra inside $\su(6)\subset\sym(12,\mathbb{R})$. The Paint group has in this case three generators and we took their linear combination with coefficients $\mathit{f}_{i=1,2,3}$. We obtain the following matrix:
\begin{equation}\label{pitturainsp}
 \mathbb{G}_{Paint} \,\ni\,X_{Paint} \,= \,\left(
\begin{array}{cc|ccc|c||cc|ccc|c}
 0 & 0 & 0 & 0 & 0 & 0 & 0 & 0 & 0 & 0 & 0 & 0 \\
 0 & 0 & 0 & 0 & 0 & 0 & 0 & 0 & 0 & 0 & 0 & 0 \\
 \hline
 0 & 0 & 0 & \mathit{f}_1 & \mathit{f}_2 & 0 & 0 & 0 & 0 & 0 & 0 & 0 \\
 0 & 0 & -\mathit{f}_1 & 0 & \mathit{f}_3 & 0 & 0 & 0 & 0 & 0 & 0 & 0 \\
 0 & 0 & -\mathit{f}_2 & -\mathit{f}_3 & 0 & 0 & 0 & 0 & 0 & 0 & 0 & 0 \\
 \hline
 0 & 0 & 0 & 0 & 0 & 0 & 0 & 0 & 0 & 0 & 0 & 0 \\
 \hline\hline
 0 & 0 & 0 & 0 & 0 & 0 & 0 & 0 & 0 & 0 & 0 & 0 \\
 0 & 0 & 0 & 0 & 0 & 0 & 0 & 0 & 0 & 0 & 0 & 0 \\
 \hline
 0 & 0 & 0 & 0 & 0 & 0 & 0 & 0 & 0 & \mathit{f}_1 & \mathit{f}_2 & 0 \\
 0 & 0 & 0 & 0 & 0 & 0 & 0 & 0 & -\mathit{f}_1 & 0 & \mathit{f}_3 & 0 \\
 0 & 0 & 0 & 0 & 0 & 0 & 0 & 0 & -\mathit{f}_2 & -\mathit{f}_3 & 0 & 0 \\
 \hline
 0 & 0 & 0 & 0 & 0 & 0 & 0 & 0 & 0 & 0 & 0 & 0 \\
\end{array}
\right)
\end{equation}
We deem that it is clear from eq.(\ref{pitturainsp}) what is the general rule. The double lines in eq.(\ref{pitturainsp}) partition the $(6+2p)\times(6+2p)$ matrix into 4 $(3+p)\times(3+p)$ blocks. The single lines separate in each of the four blocks the $p\times p$ Paint block leaving 3 additional columns and lines that surround it.
\par
It is evident form equation eq.(\ref{pitturainsp}) that the Paint subalgebra is immersed in the subalgebra $\so(3+p)\subset \uu(3+p)$ since it is block-diagonal.
\par
An interesting question is the following one: \textit{What is the normalizer subalgebra in $\uu(3+p)$ of the paint algebra?}
By this we mean:
\begin{equation}\label{lograno}
  \mathrm{N}_{\uu(3+p)}\left[\mathbb{G}_{Paint}\right]\, \equiv \,\left\{Y\in\uu(3+p)\, \mid \, \forall X\in \mathbb{G}_{Paint} \, , \, [X,Y]=0 \right\}
\end{equation}
In the case $p=3$ the normalizer is a subalgebra of dimension $10$ and its generic element has the following structure:
\begin{equation}\label{normaloide}
 Y \,= \, \left(
\begin{array}{cc|ccc|c||cc|ccc|c}
 0 & n_1 & 0 & 0 & 0 & n_2 & 2 n_4 & n_5 & 0 & 0 & 0 & n_6 \\
 -n_1 & 0 & 0 & 0 & 0 & n_3 & n_5 & 2 n_7 & 0 & 0 & 0 & n_8 \\
 \hline
 0 & 0 & 0 & 0 & 0 & 0 & 0 & 0 & 2 n_9 & 0 & 0 & 0 \\
 0 & 0 & 0 & 0 & 0 & 0 & 0 & 0 & 0 & 2 n_9 & 0 & 0 \\
 0 & 0 & 0 & 0 & 0 & 0 & 0 & 0 & 0 & 0 & 2 n_9 & 0 \\
 \hline
 -n_2 & -n_3 & 0 & 0 & 0 & 0 & n_6 & n_8 & 0 & 0 & 0 & 2 n_{10} \\
 \hline\hline
 -2 n_4 & -n_5 & 0 & 0 & 0 & -n_6 & 0 & n_1 & 0 & 0 & 0 & n_2 \\
 -n_5 & -2 n_7 & 0 & 0 & 0 & -n_8 & -n_1 & 0 & 0 & 0 & 0 & n_3 \\
 \hline
 0 & 0 & -2 n_9 & 0 & 0 & 0 & 0 & 0 & 0 & 0 & 0 & 0 \\
 0 & 0 & 0 & -2 n_9 & 0 & 0 & 0 & 0 & 0 & 0 & 0 & 0 \\
 0 & 0 & 0 & 0 & -2 n_9 & 0 & 0 & 0 & 0 & 0 & 0 & 0 \\
 \hline
 -n_6 & -n_8 & 0 & 0 & 0 & -2 n_{10} & -n_2 & -n_3 & 0 & 0 & 0 & 0 \\
\end{array}
\right)
\end{equation}
From the structure of $Y$ it is evident that it has the same form and the same number (=10) of parameters for all values of $p$.
\par
The structure of the normalizer of $\mathbb{G}_{Paint}$ in $\uu(3)$  is very important for our purposes since we need to decide whether the additional compact isometry generator of $\mathrm{L}(-1,p)$, \textit{i.e.} $\mathrm{H_L}$ derived in section \ref{kallergenerator} belongs to the $\uu(3)$ subalgebra or not.
Why such question is relevant? We know that the solvable group of L$(-1,p)$ is a subgroup of the solvable group of the Siegel half-plane. If the compact isometry subalgebra of L$(-1,p)$ were a subalgebra of $\uu(3)$ then we could state that L$(-1,p)$ is a reduction of the Siegel space. As we stressed in section
\ref{kallergenerator} the answer is that $\mathrm{H_L}$ is not contained in $\uu(3)$. Hence L$(-1,p)$ is a submanifold of the Siegel plain but it is not a Riemannian reduction of the latter.
\newpage
\section{The explicit form of the solvable generators for the $p=3$ case}
\label{trinop}
As we did in previous instances, we utilise the case $p=3$ as an illustration since, from such an example, the general structure for all values of $p$ becomes evident. The solvable Lie group element
$$\mathbb{L}_d\left(\varphi,\psi,\mathbf{x},\boldsymbol{\chi}\right) \in \mathcal{S}_{L(-1,3)} \subset \mathcal{S}^{Sieg}_{6} \subset \Sp(12,\mathbb{R}) $$
defined in eq.(\ref{sigmamapours}) of the main text is the exponential $\Sigma$-map of the solvable Lie algebra element
$X \in Solv_{L(-1,3)}$ defined in the same equation(\ref{sigmamapours}). The solvable Lie algebra element $X$ has the following explicit appearance:
\begin{eqnarray}\label{Xrep6}
& X = &\nonumber\\
& {\fontsize{6.4}{6.5} \left(
\begin{array}{cccccc|cccccc}
 \frac{3 \psi }{2} & 2 x^1 & -4 x^2 & -4 x^3 & -4 x^4 & 8 x^5 & 0 & 0 & 0 & 0 & 0 & 0 \\
 0 & \frac{\psi }{2}-\frac{\varphi }{\sqrt{6}} &\sqrt{2}  \chi^1 &\sqrt{2}  \chi^2 &\sqrt{2}  \chi^3 & 0 & 0 & 8 x^5 & 4
   x^2 & 4 x^3 & 4 x^4 & 2 x^1 \\
 0 & 0 & \frac{\varphi }{2 \sqrt{6}}+\frac{\psi }{2} & 0 & 0 & 2 \sqrt{2}\chi^1 & 0 & 4 x^2 & -2 x^1 & 0 & 0 & 0 \\
 0 & 0 & 0 & \frac{\varphi }{2 \sqrt{6}}+\frac{\psi }{2} & 0 & 2 \sqrt{2} \chi^2 & 0 & 4 x^3 & 0 & -2 x^1 & 0 & 0 \\
 0 & 0 & 0 & 0 & \frac{\varphi }{2 \sqrt{6}}+\frac{\psi }{2} & 2 \sqrt{2} \chi^3 & 0 & 4 x^4 & 0 & 0 & -2 x^1 & 0 \\
 0 & 0 & 0 & 0 & 0 & \sqrt{\frac{2}{3}} \varphi +\frac{\psi }{2} & 0 & 2 x^1 & 0 & 0 & 0 & 0 \\
 \hline
 0 & 0 & 0 & 0 & 0 & 0 & -\frac{3 \psi }{2} & 0 & 0 & 0 & 0 & 0 \\
 0 & 0 & 0 & 0 & 0 & 0 & -2 x^1 & \frac{\varphi }{\sqrt{6}}-\frac{\psi }{2} & 0 & 0 & 0 & 0 \\
 0 & 0 & 0 & 0 & 0 & 0 & 4 x^2 & -\sqrt{2} \chi^1 & -\frac{\varphi }{2 \sqrt{6}}-\frac{\psi }{2} & 0 & 0 & 0 \\
 0 & 0 & 0 & 0 & 0 & 0 & 4 x^3 & -\sqrt{2}\chi^2 & 0 & -\frac{\varphi }{2 \sqrt{6}}-\frac{\psi }{2} & 0 & 0 \\
 0 & 0 & 0 & 0 & 0 & 0 & 4 x^4 & -\sqrt{2} \chi^3 & 0 & 0 & -\frac{\varphi }{2 \sqrt{6}}-\frac{\psi }{2} & 0 \\
 0 & 0 & 0 & 0 & 0 & 0 & -8 x^5 & 0 & -2\sqrt{2} \chi^1 & -2\sqrt{2}\chi^2 & -2\sqrt{2}\chi^3 & -\sqrt{\frac{2}{3}}
   \varphi -\frac{\psi }{2} \\
\end{array}
\right)}&\nonumber\\
\end{eqnarray}
Replacing in eq.(\ref{Xrep6}) the coordinate transformation presented in eq.(\ref{luchang}), one obtains:
\begin{eqnarray}
\label{Xr6w}
 X=\hat{X}(\mathbf{w}) &= &\left(
                  \begin{array}{c|c}
                    A_X(\mathbf{w}) & B_X(\mathbf{w}) \\
                    \hline
                    \mathbf{0} & - A_X^T(\mathbf{w}) \\
                  \end{array}
                \right)
\end{eqnarray}
where
\begin{equation}\label{AXeq}
  A_X \, = \,\left(
\begin{array}{cccccc}
 \frac{1}{2} \left(3 w^1-w^2\right) & -2 w^3 & -4 w^{5,1} & -4 w^{5,2} & -4 w^{5,3} & 16 w^4 \\
 0 & \frac{1}{2} \left(w^1+w^2\right) & \sqrt{2} e^{w^2} w^{6,1} &\sqrt{2} e^{w^2} w^{6,2} & \sqrt{2} e^{w^2} w^{6,3} & 0 \\
 0 & 0 & \frac{1}{2} \left(w^1-w^2\right) & 0 & 0 & 2\sqrt{2}\,e^{w^2} w^{6,1} \\
 0 & 0 & 0 & \frac{1}{2} \left(w^1-w^2\right) & 0 & 2\sqrt{2}\, e^{w^2} w^{6,2} \\
 0 & 0 & 0 & 0 & \frac{1}{2} \left(w^1-w^2\right) & 2\sqrt{2}\,e^{w^2} w^{6,3} \\
 0 & 0 & 0 & 0 & 0 & \frac{1}{2} \left(w^1-3 w^2\right) \\
\end{array}
\right)
\end{equation}
and
\begin{equation}\label{BXeq}
  B_X \, = \,\left(
\begin{array}{cccccc}
 0 & 0 & 0 & 0 & 0 & 0 \\
 0 & 16 w^4 & 4 w^{5,1} & 4 w^{5,2} & 4 w^{5,3} & -2 w^3 \\
 0 & 4 w^{5,1} & 2 w^3 & 0 & 0 & 0 \\
 0 & 4 w^{5,2} & 0 & 2 w^3 & 0 & 0 \\
 0 & 4 w^{5,3} & 0 & 0 & 2 w^3 & 0 \\
 0 & -2 w^3 & 0 & 0 & 0 & 0 \\
\end{array}
\right)
\end{equation}
The next step, in order to obtain the explicit form of the generators in the solvable coordinate basis $\mathbf{w}$ of CV manifolds, is that of expanding the matrix $\hat{X}(\mathbf{w})$ to first order in the $\mathbf{w}$-coordinates:
\begin{equation}\label{gronaldo}
  \hat{X}(\mathbf{w})\, = \, \mathbf{w}\cdot \mathbf{T} \, + \,\mathcal{O}(\mathbf{w}^2)
\end{equation}
In this way one obtains
\begin{eqnarray}
\label{aenadumgenetrix}
  \mathbf{w}\cdot \mathbf{T}&=& w^1\, T_1 + w^2\, T_2 + w^3\, T_3+ w^4 \,T_4 + w^{5,\ir}\, T_{5,\ir} + w^{6,\ir} \,T_{6,\ir} \nonumber\\
  \mathbf{w}\cdot \mathbf{T}&=& \left(
                  \begin{array}{c|c}
                    \mathcal{A}(\mathbf{w}) & \mathcal{B}(\mathbf{w}) \\
                    \hline
                    \mathbf{0} & - \mathcal{A}^T(\mathbf{w}) \\
                  \end{array}
                \right)
\end{eqnarray}
where
\begin{equation}\label{Afin}
  \mathcal{A}(\mathbf{w})\,=\,\left(
\begin{array}{cccccc}
 \frac{1}{2} \left(3 w^1-w^2\right) & -2 w^3 & -4 w^{5,1} & -4 w^{5,2} & -4 w^{5,3} & 16 w^4 \\
 0 & \frac{1}{2} \left(w^1+w^2\right) & \sqrt{2}w^{6,1} & \sqrt{2} w^{6,2} & \sqrt{2} w^{6,3} & 0 \\
 0 & 0 & \frac{1}{2} \left(w^1-w^2\right) & 0 & 0 & 2\sqrt{2} \, w^{6,1} \\
 0 & 0 & 0 & \frac{1}{2} \left(w^1-w^2\right) & 0 & 2\sqrt{2} \, w^{6,2} \\
 0 & 0 & 0 & 0 & \frac{1}{2} \left(w^1-w^2\right) & 2\sqrt{2} \,w^{6,3} \\
 0 & 0 & 0 & 0 & 0 & \frac{1}{2} \left(w^1-3 w^2\right) \\
\end{array}
\right)
\end{equation}
and
\begin{equation}\label{Bfin}
  \mathcal{B}(\mathbf{w})\,=\,\left(
\begin{array}{cccccc}
 0 & 0 & 0 & 0 & 0 & 0 \\
 0 & 16 w^4 & 4 w^{5,1} & 4 w^{5,2} & 4 w^{5,3} & -2 w^3 \\
 0 & 4 w^{5,1} & 2 w^3 & 0 & 0 & 0 \\
 0 & 4 w^{5,2} & 0 & 2 w^3 & 0 & 0 \\
 0 & 4 w^{5,3} & 0 & 0 & 2 w^3 & 0 \\
 0 & -2 w^3 & 0 & 0 & 0 & 0 \\
\end{array}
\right)
\end{equation}
Extracting the generators $T_A =\{T_{1},\ldots, T_4,T_{5,i},T_{6,i}\}$
from their definition provided by eq.s(\ref{aenadumgenetrix}-\ref{Bfin}) one can verify that they satisfy a closed solvable Lie Algebra with structure constants identical to those defined by the Maurer Cartan eq.s (\ref{maurocartus})
\newpage
\section{Short Summary of the Homogeneous Special Geometry Classification}
\label{classmates}
In this appendix, relying on the old literature on the classification of Supergravity Special Geometries
\cite{productproof,SKGaggio3,SKGaggio2,SKGaggio1,specHomgeoA2,specHomgeoA1,vandersuppa,Ferrara:1989py,deWit:1995tf,Lauria:2020rhc}
and on its review in the book \cite{Freedman:2012zz} we show how the identification of the solvable group for the L$(-1,p)$ and L$(0,p)$ spaces is rooted in the structure of the corresponding cubic polynomials pinpointing the possible extension of the same mechanism to other cases of homogeneous special geometries.
\par
First we recall the nomenclature. The manifolds L$(q,P)$ are defined by a cubic polynomial
\begin{equation}
  {\cal C}_{\ial\jbe\kga}\,h^\ial\,h^\jbe\,h^\kga = 3\left\{ h^1\,\big(h^2\big)^2 -h^1\,
\big(h^\mua\big)^2 -h^2\,\big(h^\ir\big)^2 +\Red{\gamma_{\mua \ir\js}}\,h^\mua\,
h^\ir\,h^\js \right\}
 \label{cubicpolLqP}
\end{equation}
where the index $\mua $ runs over $q+1$ values, and the variables $h^\ir$ form $P$ irreps of the real Clifford algebra ${\cal C}(q+1,0)$, with gamma-matrices $\gamma _{\mua \ir\js}$. Thus $\ir$ runs over $(P+\dot P){\cal D}_{q+1}$ values where ${\cal D}_{q+1}$ is given in table \ref{tbl:Dq1}.
\begin{table}[htb]
  \centering
 \begin{tabular}{||c|c|c||}\hline
$q$ &${\cal C}(q+1,0)$& ${\cal D}_{q+1}$ \\ \hline &&\\[-3mm]
$-1$ &$\Rbar$    &1         \\
0    &$\Rbar\oplus \Rbar $&1   \\
1    &$\Rbar(2)$ &2           \\
2    &$\Cbar(2)$ &4            \\
3    &$\Hbar(2)$ &8             \\
 4   &$\Hbar (2)\oplus \Hbar (2)$&8       \\
5    &$\Hbar(4)$ &16   \\
6    &$\Cbar(8)$ &16    \\
7    &$\Rbar(16)$&16\\
$n+7$& $\Rbar(16)\otimes{\cal C}(n,0)$&16 ${\cal D}_n$ \\[1mm]
\hline
\end{tabular}
  \caption{The values of ${\cal D}_{q+1}$ }\label{tbl:Dq1}
\end{table}
\par
Note that ${\cal D}_{q+1}$ is the same for $q=4m-1$ and $q=4m$. The $\gamma $-matrices for $q=4m$ can be obtained from those for $q=4m-1$, adding $\gamma _{4m+1}\equiv \gamma _1\gamma _2\ldots\gamma _{4m}$, which is then also a real matrix of the same size that squares to 1. In the case $m=0$ considered before, the $h^\mua $ are absent for $q=-1$, but for $q=0$, the $\gamma _1$ is just a unit matrix.
\par
This leads to $n=3+q+(P+\dot P){\cal D}_{q+1}$. We restrict now for $q=4m$ to $\dot P=0$. Thus the $n$ value for $q=4m$ is just one higher than the one for $q=4m-1$, corresponding to adding the $h^\mu $  for $\mu =4m+1$.
\par
Note that also for $m=1$ and $m=2$ (as for $m=0$), the L$(4m,1)$ is a symmetric space as well for the real, complex and quaternionic version. Indeed, comparing
 L$(4m-1,P)$ with L$(4m,P)$, the only symmetric space cases (for the K\"{a}hler version of the special manifold) are $\mathrm{L}(0,p) \, = \, \frac{\mathrm{SO}(2,2+p)}{\mathrm{SO}(2) \times \mathrm{SO}(2+p)}$ (case $m=0$),  $\mathrm{L(4,1)}= \frac{\mathrm{SO^*(12)}}{\mathrm{SU(6)\otimes U(1)}}$ (case $m=1$), and $\mathrm{L(8,1)}= \frac{\mathrm{E_{7(-25)}}}{\mathrm{E_{6(-78)} \otimes U(1)}}$ (case $m=4$).
\par
We also see that the Paint Groups $\mathrm{G_{Paint} }\, = \, {\cal S}_{4m-1}(P)={\cal S}_{4m}(P,0)$ are the same\footnote{We apologize with our reader for another unavoidable source of notational confusion in the comparison of  the results from classical Supergravity literature on the classification of Special Geometries (see in particular \cite{titsusataku}) and the results established in the recent literature on CaNNs \cite{pgtstheory,TSnaviga,naviga,tassellandum,axialbeltra,geotermico,secondtemperature,terzatemperatura}.
In the latter we use the letter $\mathcal{S}_\mathcal{M}\equiv \exp[Solv_\mathcal{M}]$ to denote the solvable Lie Group, metric equivalent to the considered non-compact manifold $\mathcal{M}$ while in \cite{titsusataku} and in \cite{productproof,SKGaggio3,SKGaggio2,SKGaggio1,specHomgeoA2,specHomgeoA1,vandersuppa,Ferrara:1989py,deWit:1995tf,Lauria:2020rhc}
${\cal S}_{q}(P)$ was utilized to denote the Paint Group. }. For $m$ even this is $\SO(P)$ and for $m$ odd, this is $\USp(2P)$
\par
It is interesting to note that the mentioned cases L$(4,1)$ and L$(8,1)$ are two members, together with other two non-compact symmetric spaces, in a short, finite, Tits Satake universality class that is explicitly displayed in table 5.4 at page 235 of the book \cite{advancio}. The 4 member TS class is recalled in table \ref{cirimello}.
\begin{table}[htb]
\begin{center}
$$\label{uniclastscorta}
  \begin{array}{||c||c||}
  \hline
  \hline
  \text{TS submanifold}&\text{Class Members}\\
  \hline
  \hline
  \frac{\Sp(6,\mathbb{R})}{\mathrm{SU(3)}\times \mathrm{U(1)}}& \begin{array}{c|c|c}
  \null & \null & \null \\
  \text{Symm. Space} & \text{Paint Group} & \text{subPaint Group}\\
  \null & \null & \null \\
  \hline
  \null & \null & \null \\
  \frac{\Sp(6,\mathbb{R})}{\mathrm{SU(3)}\times \mathrm{U(1)}} & \mathbf{1} & \mathbf{1}\\
  \null & \null & \null \\
  \hline
  \null & \null & \null \\
  \frac{\mathrm{SU(3,3)}}{\mathrm{SU(3)}\times\mathrm{SU(3)}\times \mathrm{U(1)}} & \mathrm{SO(2)} \times \mathrm{SO(2)} & \mathbf{1} \\
  \null & \null & \null \\
  \hline
  \null & \null & \null \\
  \frac{\mathrm{SO^\star(12)}}{\mathrm{SU(6)}\times \mathrm{U(1)}} & \mathrm{SO(3)} \times \mathrm{SO(3)}\times \mathrm{SO(3)} & \mathrm{SO(3)}_d \\
  \null & \null & \null \\
   \hline
  \null & \null & \null \\
  \frac{\mathrm{E_{7(-25)}}}{\mathrm{E_{6(-78)}\times \mathrm{U(1)}}} & \mathrm{SO(8)}  & \mathrm{G_{2(-14)}} \\
  \null & \null & \null \\
  \end{array}
  \\
  \hline
  \hline
  \end{array}
$$
\end{center}
\caption{\label{cirimello} The $4$-member Tits Satake universality class that includes the special K\"ahler manifolds L(4,1) and L(8,1)}
\end{table}
\par
Obviously such a short class, that mathematically is very interesting since it involves exceptional Lie Algebras, is too short to allow the construction of an ordinary CaNNs. Yet it might be a starting point for a convolutional CaNN, in the spirit of the forthcoming paper \cite{convopietro}. Whether this makes sense and it might be done, crucially depends on a study of the subTits Satake manifold of the Siegel halfplane of genus $g=3$ and of the nesting of representations of one Paint Group inside the next one. In any case it is a suggestion not to be dismissed a priori. Exceptional Lie algebras were considered a mathematical curiosity until the advent of Supergravity and Superstrings that brought them to the forefront. It would be interesting to check whether they play an unexpected relevant role also in Machine Learning.
\par
Coming back to the cubic form ${\cal C}$ for $q=0$ it can be simplified so that it looks very similar to the one for $q=-1$, while we did not find such a simplification for $q=4m,\ m>0$. Indeed we have
\begin{align*}
  \mathrm{L}(-1,P)\ : \ \frac{1}{3} {\cal C}&= h^1 (h^2)^2 -h^2(h^\ir)^2\,,\nonumber\\
  \mathrm{L}(0,P)\ : \  \frac{1}{3} {\cal C}&= h^1 (h^2)^2 -h^1 (h^3)^2-h^2(h^\ir)^2=h^3(h^\ir)^2 \nonumber\\
  &=h^1(h^2+h^3)(h^2-h^3)(h^2+h^3) -(h^2-h^3)(h^\ir)^2\,.
\end{align*}
This algebraic identity makes the connection between the two solvable groups already clear.
\subsection{Comparison}
\label{kumpare}
The figures in \cite[Fig.1]{vandersuppa} and  \cite[Table 11]{deWit:1995tf} make it easy to compare with e.g. (\ref{lagradisca}) in the main body of the article.  We take the axis $x=-2a_2$ and $y=a_1\sqrt{\frac{3}{2}}$ in (\ref{lagradisca}), such that the gradings of the $E$ symmetries are
\begin{equation}
  E_{\chi^r}: \ (0,1)\,,\qquad E_{x^1}:\ ((1,-\frac{2}{3})\,,\qquad E_{x^{r+1}}:\ (1,\frac{1}{3})\,,\qquad x^n:\ (1,\frac{4}{3})\,.
 \label{coordinatesE}
\end{equation}
This is represented in table \ref{tbl:isohomvsk}, which is similar to the mentioned figures in the old papers. We kept $a_M$ as the additional symmetry beyond the solvable algebra. It is the generator $L_-$ explicitly constructed inside $\sym(6+2p,\mathbb{R})$ in  section \ref{sp6extragen} and inside
$\so(2,2+p)$ in section \ref{pseudortho}.
\begin{table}[ht]\caption{Isometries of non-symmetric
homogeneous special K\"{a}hler manifolds, and explicitly for L$(-1,P)$}
\label{tbl:isohomvsk}
\begin{center}
\setlength{\unitlength}{0.6mm}
\begin{picture}(200,70)
\put(20,30){\line(1,0){110}}
\put(75,0){\line(0,1){70}}
\put(75,30){\circle*{2}}
\put(75,30){\circle{5}}
\put(75,60.6){\circle*{2}}
\multiput(103.9,9.6)(0,30.6){3}{\circle*{2}}
\put(65,33){\makebox(0,0)[bl]{$a_2$}}
\put(65,63.6){\makebox(0,0)[bl]{$\chi^r$}}
\put(77.4,21){\makebox(0,0)[bl]{$a_1 $}}
\put(106.3,6.6){\makebox(0,0)[bl]{$x^1$}}
\put(106.3,37.2){\makebox(0,0)[bl]{$x^{r+1}$}}
\put(106.3,67.8){\makebox(0,0)[bl]{$x^n$}}
\put(46.1,50.4){\makebox(0,0){\circle*{2}} }
\put(43.7,47.4){\makebox(0,0)[br]{$a_M$}}
\end{picture}
\end{center}  \end{table}

\end{document}